\documentclass[twocolumn]{aastex62}
\usepackage{epstopdf}
\usepackage{graphicx}
\usepackage{natbib}
\usepackage{float}
\usepackage[utf8]{inputenc}
\usepackage{amsmath}
\usepackage{bm}
\usepackage[title,toc,titletoc]{appendix}
\usepackage{comment}
\usepackage{academicons}

\definecolor{orcidlogocol}{HTML}{A6CE39}

\newcommand{\singlebe}{\texttt{SINGLEBE}}

\newcommand{\emcee}{\texttt{emcee}}
\newcommand{\hdust}{\texttt{HDUST}}

\begin{document}

\title{The Be Star 66 Ophiuchi: 60 Years of Disk Evolution}

\author[0000-0003-0989-2941]{K. C. Marr}
\affil{Department of Physics and Astronomy, University of Western Ontario, London, ON N6A 3K7, Canada}

\author[0000-0001-9900-1000]{C. E. Jones}
\affiliation{Department of Physics and Astronomy, University of Western Ontario, London, ON N6A 3K7, Canada}

\author[0000-0002-9369-574X]{A. C. Carciofi}
\affiliation{Instituto de Astronomia, Geofísica e Ciências Atmosféricas, Universidade de São Paulo, Rua do Matão 1226, Cidade Universitária, 05508-900 São Paulo, SP, Brazil}

\author[0000-0002-2490-1562]{A. C. Rubio}
\affiliation{Instituto de Astronomia, Geofísica e Ciências Atmosféricas, Universidade de São Paulo, Rua do Matão 1226, Cidade Universitária, 05508-900 São Paulo, SP, Brazil}

\author[0000-0002-7851-4242]{B. C. Mota}
\affiliation{Instituto de Astronomia, Geofísica e Ciências Atmosféricas, Universidade de São Paulo, Rua do Matão 1226, Cidade Universitária, 05508-900 São Paulo, SP, Brazil}

\author[0000-0003-3682-6691]{M. R. Ghoreyshi}
\affiliation{Department of Physics and Astronomy, University of Western Ontario, London, ON N6A 3K7, Canada}

\author[0000-0001-5576-6669]{D. W. Hatfield}
\affiliation{Department of Physics and Astronomy, University of Western Ontario, London, ON N6A 3K7, Canada}

\author[0000-0002-1427-3662]{L. R. R\'imulo}
\affiliation{Instituto de Astronomia, Geofísica e Ciências Atmosféricas, Universidade de São Paulo, Rua do Matão 1226, Cidade Universitária, 05508-900 São Paulo, SP, Brazil}

\begin{abstract}
We use a time-dependent hydrodynamic code and a non-LTE Monte Carlo code to model disk dissipation for the Be star 66 Ophiuchi. We compiled 63 years of observations from 1957 to 2020 to encompass the complete history of the growth and subsequent dissipation of the star's disk. Our models are constrained by new and archival photometry, spectroscopy and polarization observations, allowing us to model the disk dissipation event. Using Markov chain Monte Carlo methods, we find 66 Oph is consistent with standard B2Ve stellar properties. We computed a grid of $61568$ Be star disk models to constrain the density profile of the disk before dissipation using observations of the H$\alpha$ line profile and SED. We find at the onset of dissipation the disk has a base density of $2.5\times10^{-11}\ \rm{g\ cm^{-3}}$ with a radial power-law index of $n=2.6$. Our models indicate that after $21$ years of disk dissipation 66 Oph's outer disk remained present and bright in the radio. We find an isothermal disk with constant viscosity with an $\alpha = 0.4$ and an outer disk radius of $\sim115$ stellar radii best reproduces the rate of 66 Oph's disk dissipation. We determined the interstellar polarization in the direction of the star in the V-band is $p=0.63 \pm 0.02\%$ with a polarization position angle of $\theta_{IS}\approx85.7 \pm 0.7^\circ$. Using the Stokes QU diagram, we find the intrinsic polarization position angle of 66 Oph's disk is $\theta_{int}\approx98 \pm 3^\circ$.
\end{abstract}


\section{Introduction}\label{sec:intro}
Classical B-emission (Be) stars are rapidly rotating, main sequence, B-type stars that form outwardly diffusing disks of gas that have been ejected from the stellar surface. These geometrically thin disks are comprised of almost fully-ionized hydrogen gas and rotate around the stellar equator in Keplerian fashion \citep{rivi2013}. The rapid rotation of these stars \citep{slet1982} is believed to lead to the disk formation, which is likely assisted by non-radial pulsations \citep{rivi2003, baade2016}. The disks are characterized by infrared and radio excesses in the star's spectrum, polarized light, and the presence of hydrogen series emission lines \citep{rivi2013}.

Many Be stars show variability over timescales ranging from minutes \citep{gora2008} to decades \citep{okaz1997}. The rapid variability observed from these systems is commonly associated with localized mass ejections \citep{balo1990}, $\beta$-Cephei type pulsations \citep{balo1991} and non-radial pulsations \citep{baade1982, rivi2003}. Some disks exhibit cycles of growth and dissipation, which are dependent on the viscosity of the gas as outlined in the viscous decretion disk (VDD) model \citep{lee91, papa92, klem15}. This type of variability occurs typically over periods of decades \citep[e.g.][]{wisn2010}.

66 Ophiuchi (HR 6712, HD 164284) is a bright ($m_v\sim4.8~\rm{mag}$), multiple star system \citep{stefl2004} containing a classical Be star of spectral type B2Ve \citep{floq2002}, at a distance of $\sim$199.6 pc\footnote{Based on \textit{Hipparcos} parallaxes \citep{vanl2007}}. Since 1957, observations indicate that 66~Oph built a large disk which it subsequently lost over a period of dissipation that started around 1990 and finished about 20 years later. These events make 66 Oph an ideal system for studying the physics of disk dissipation.

Many observational campaigns have focused on this star because of the variability of its disk. In 1957, the transition of the H$\alpha$ line into emission signalled the onset of disk formation. \citep{rako1958}. \citet{grady1987}, \citet{percy1992} and \citet{percy2001} found that metal lines at UV wavelengths showed evidence of recurrent episodic mass-loss from 1982 to 1985, and again from 1989 to 1999 using UBV photometry. \citet{hanu1996} reported that the ratio of the violet (V) and red (R) peaks of the H$\alpha$ profile began to vary in 1988, and continued through 1995 when it had a period of five years. \citet{stefl2004} showed that the V/R ratio was constant (i.e. V = R) at the onset of dissipation, suggesting the disk was axisymmetric. \citet{floq2002} gave a detailed history of the pulsation period of 66 Oph before the period of disk dissipation, which they claim began in 1992. Since then, there has been a slow decline of the H$\alpha$ emission strength with a transition to an absorption line in 2010 \citep{sabo2017}. The line profile has remained unchanged since then.

In this study, we investigate the evolution of 66 Oph's disk at the onset and throughout the dissipation period. We characterized the physical state of the disk before dissipation and the dynamics of the disk during dissipation. This includes identifying the density structure and radial extent of the disk, as well as the viscosity and temperature distribution. This was accomplished by finding the best model to reproduce observations of the spectral energy distribution (SED), H$\alpha$ spectral line, V-band polarization and V-band photometry during dissipation. By including observations from previous works in addition to presenting new observations, our unique data set allows us to model the complete dissipation of 66 Oph's disk for the first time.

The models presented in this work were created following similar methods to \citet{haub2012}, \citet{carc2012}, \citet{rimu2018} and \citet{ghor2018}. The 1D dynamical code \singlebe\ was used to compute the disk surface density of an axisymmetric disk as a function of time, given a disk viscosity. The resulting disk density distributions were then input to the 3D radiative transfer code \hdust\ to calculate the emergent spectrum. In Section~\ref{sec:observations}, we describe the observations compiled over the period of 1957 to 2020. Section~\ref{sec:stellarparams} gives a detailed description of our method to determine the parameters of the star. Section~\ref{sec:results} describes our modelling routines along with the results of the modelling. Section~\ref{sec:discussion} provides a discussion and summary of our work including a comparison to other findings in the literature.

\section{Observations}\label{sec:observations}

\begin{figure*}
	\centering
	\makebox[\textwidth][c]{\includegraphics[width = 1\textwidth]{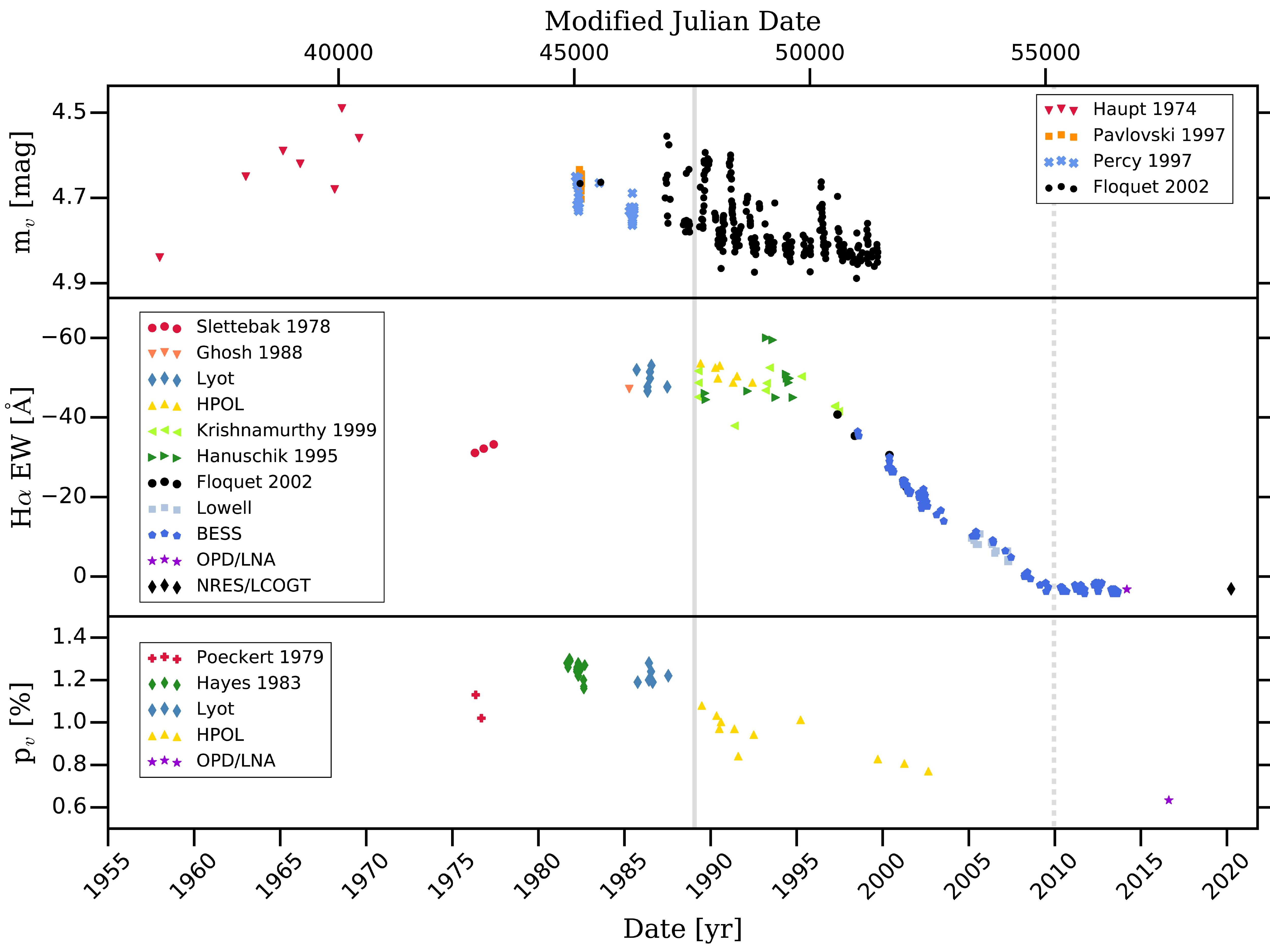}}
	\caption{Observations of 66 Oph from 1957 to 2020. \textit{Top:} V-band photometry \textit{Middle:} H$\alpha$ EW \textit{Bottom:} V-band polarization. The sources of the data are listed in the legend; those without dates are previously unpublished. The solid grey vertical line indicates the time for the onset of disk dissipation and the dashed grey vertical line indicates when the H$\alpha$ line transitioned to absorption. The solid and dashed grey lines also approximately correspond to the periods of observation for IRAS and WISE, respectively.}
	\label{fig:allplots}
\end{figure*}

We compiled observations from a variety of different sources to investigate the disk's evolution over many decades. Figure~\ref{fig:allplots} shows the V-magnitude photometry, H$\alpha$ equivalent width (henceforth, EW) and continuum V-band polarization from the onset of disk growth and subsequent dissipation from 1957 to 2020. We adopt the convention that negative EW means flux above the continuum.

Most of the observations used in this work were from the literature. We used V-magnitude photometry from the works of \citet{haupt1974}, \citet{pavl1997}, \citet{percy1997}, and \citet{floq2002}. We also acquired H$\alpha$ EW's from the works of \citet{slet1978}, \citet{ghos1988}, \citet{kris1999}, \citet{hanu1995}, \citet{floq2002}, and from the Be Star Spectra Database (BeSS)\footnote{http://basebe.obspm.fr/basebe/}. Reported values of the H$\alpha$ EW and V-band polarization were acquired from the archive for the Lyot Spectropolarimeter\footnote{http://www.sal.wisc.edu/PBO/LYOT/} and the Halfwave Spectropolarimeter (HPOL) at the University of Wisconsin-Madison Pine Bluff Observatory, which were reduced in the work of \citet{drap2014}. We also used observations of the V-band polarization published by \citet{poeck1979} and \citet{hayes1983}.

A number of previously unreported observations of 66 Oph are also presented. We used measurements of the H$\alpha$ EW determined from observations of the H$\alpha$ spectra made using the fibre-fed échelle spectrograph attached to the 1.1 meter John S. Hall telescope at the Lowell Observatory in Flagstaff, Arizona. Observations from this instrument were obtained between 2005 and 2007, with a resolving power of $R = 10000$. The reduction process of these observations follows that previously described in \citet{jones2017}.

Our models are constrained by observations of the H$\alpha$ EW and V-band polarization, acquired respectively using the MUSICOS spectrograph and the IAGPOL polarimeter at the Pico dos Dias Observatory (OPD), operated by the National Astrophysical Laboratory of Brazil (LNA) in Minas Gerais, Brazil. These observations were reduced with packages developed by the B\textsc{eacon} group\footnote{http://beacon.iag.usp.br/}, and described in \citet{maga1984, maga1996} and \citet{carc2007}.

The most recent H$\alpha$ EW observation was obtained by the NRES spectrograph at the Las Cumbres Observatory LCOGT network. Details about the instrument and reduction process of this observation can be found in \citet{brown2013}. Since this observation was acquired while 66 Oph was diskless, it was used for comparison to our diskless models. This observation is later presented in Subsection \ref{subsec:results_2}.

Over a 63 year period, the V-band photometry, $m_{\nu}$ was observed to range between $4.5 < m_{\nu} < 4.9~\rm{mag}$. During dissipation, the episodic variability continued with nine outbursts of between $1991$ and $2008$, while the overall brightness asymptotically dimmed.

As the V-band photometry available to us is sparse during the disk building phase, the variation of the light could not be used to model the evolution of the disk as was done in the studies of \citet{ghor2018} and \citet{rimu2018}. Photometric observations in other bands (UV, IR, etc.) are also sparse, with only snapshots available. Here, we use the V-band photometry along with observations of the entire SED, the H$\alpha$ line profile and the percentage of polarized light over time.

The vertical, solid grey line in Figure~\ref{fig:allplots} indicates the onset of dissipation (and is further discussed in Subsection \ref{subsec:results_1}). As the dissipation event begins, the continuum flux drops and the inner disk re-accretes causing the H$\alpha$ EW to increase. This is seen in Figure~\ref{fig:allplots} around 1995 as shown by the observations from \citet{kris1999} and \citet{hanu1995}. After this time, the EW began to steadily decrease until the line went into absorption in 2010 (indicated by the vertical, dashed grey line). Our most recent observation, obtained by NRES/LCOGT in March 2020, indicates that 66 Oph continues to be diskless.

The bottom panel of Figure~\ref{fig:allplots} shows the change in the observed polarization, $p_{v}$, with time for 66 Oph. Since 1980, the percent polarization slowly decayed until it approached the interstellar base level. From the base level polarimetric observations obtained by OPD/LNA in 2017 (listed in Table \ref{tab:polarization}), the V-band interstellar polarization is $\sim0.63\%$ with a polarization position angle of $\theta \approx 85.7^{\circ}$. This average value was subtracted from each of the observations using the scheme outlined by \citet{quir1997}. The intrinsic polarization of 66 Oph obtained is discussed further in Section \ref{subsubsec:results_2_2}.

\begin{figure}
	\centering
	\makebox[\columnwidth][c]{\includegraphics[width = 1\columnwidth]{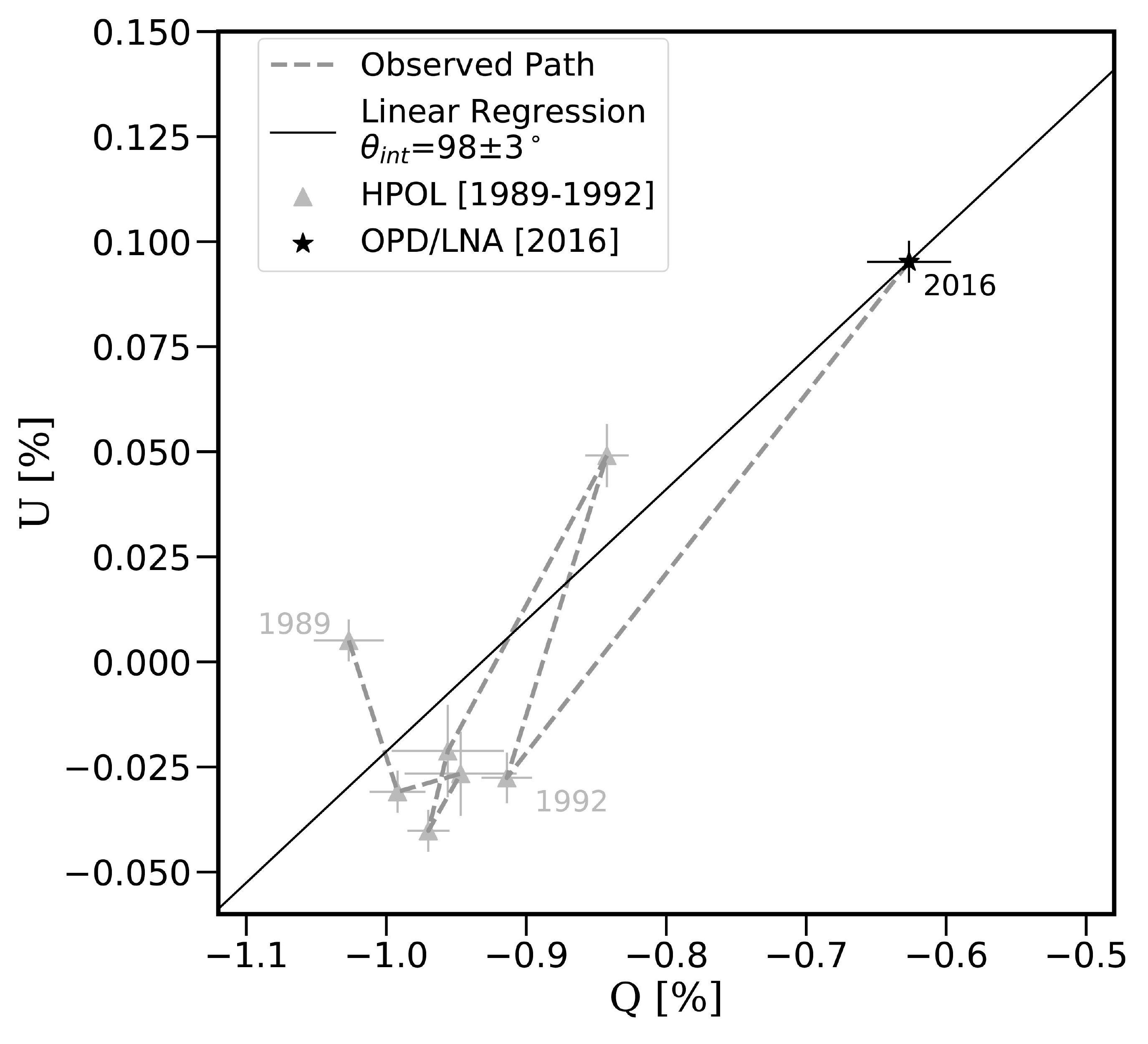}}
	\caption{V-band polarization of 66 Oph from HPOL and OPD/LNA plotted on a Stokes QU diagram. The linear regression was used to determine the intrinsic polarization position angle.}
	\label{fig:quplot}
\end{figure}

\begin{table}
	\centering
	\begin{tabular}{cclll}
		\hline
		\hline
		Modified Julian Date & Filter & P[$\%$] & $\theta$[$^\circ$] &\\
		\hline
		57615.04 & I & 0.54 $\pm$ 0.03 & 85.1 $\pm$ 1.6 &  \\  
		57615.05 & R & 0.61 $\pm$ 0.01 & 86.3 $\pm$ 0.5 &  \\  
		57615.05 & V & 0.63 $\pm$ 0.02 & 85.7 $\pm$ 0.7 &  \\ 
		57615.06 & B & 0.60 $\pm$ 0.01 & 83.9 $\pm$ 0.2 &  \\ 
		57623.12 & I & 0.51 $\pm$ 0.01 & 86.6 $\pm$ 0.5 &  \\
		\hline
		\hline
	\end{tabular}
	\caption{OPD/LNA observations of 66 Oph used in this work as an estimate of the interstellar polarization.}
	\label{tab:polarization}
\end{table}

The star's SED was also used to refine our models. We selected $112$ observations of the UV portion of 66 Oph's SED observed by the International Ultraviolet Explorer (IUE) telescope from the INES database \citep{wams2001} following the selection procedure described by \citet{ferr2012}. The data selected were observed using the large aperture and high dispersion modes on the LWR and SWP cameras to ensure proper flux calibration and high spectral resolution. The observations were obtained over the period of March 1980 to September 1995, during which the observed flux showed no significant changes. We chose to remove the IUE data beyond $0.3~\micron$ due to instrumental limitations which cause significant uncertainty \citep{wams2001}.

Visible and IR wavelength observations were obtained from the CDS Portal application from the Universit\'e de Strasbourg. Observations at these wavelengths were acquired while the disk was present (e.g. IRAS in 1989) and absent (e.g. AKARI in 2006-07 and WISE in 2010), providing upper and lower limits for modelling the SED. In addition to indicating the onset of dissipation and transition to absorption, the grey lines in Figure \ref{fig:allplots} also lie at the approximate dates of observation of IRAS (solid grey line) and WISE (dashed grey line) data. We also collected radio observations from VLA and ATCA \citep{appa1990, clark1998} and JVLA \citep{klem19}. These observations are presented alongside our models in Subsection \ref{subsec:results_1}.

\subsection{The Interstellar Polarization}\label{sec:observations_pol}

Figure \ref{fig:quplot} shows a Stokes QU diagram of the V-band polarization obtained by HPOL and by OPD/LNA. The observed path on the QU diagram during the disk dissipation phase is upward and to the right. As discussed by \citep{drap2014} and more recently by \citep{ghor2021}, the process of formation and/or dissipation of a Be star disk is associated with a linear trend in the QU diagram, during which the polarization level rises or lowers in response to changes in the disk density, but the polarization angle remains steady. By fitting a linear regression to the HPOL observations, and fixing it to the assumed diskless 2016 observation, we determined the polarization position angle of the disk to be $\sim98\pm3^{\circ}$. This was also confirmed by fitting a Serkowski law \citep{serk1973} to the OPD/LNA observations in Table \ref{tab:polarization}, using
\begin{equation}
    P(\lambda)/P(\lambda_{max}) = \rm{exp}[-1.15~\rm{ln}^2(\lambda_{max}/\lambda)].
    \label{eqn:serklaw}
\end{equation}
\noindent
We determined this law to best fit the observations in Table \ref{tab:polarization} when $P(\lambda_{max}) = 0.62 \pm 0.02\%$ at $0.61 \pm 0.04~\micron$. Subtracting this spectrum from the HPOL observation, we accurately modelled the polarization of the disk (shown later in Subsection \ref{subsubsec:results_2_2}). From this spectrum, the wavelength averaged polarization position angle of the disk was computed to be $96 \pm 4^\circ$, in perfect agreement with the estimate made using the QU diagram.

\section{Stellar Parameters}\label{sec:stellarparams}

The use of Markov chain Monte Carlo (MCMC) routines along with Bayesian statistics has recently found success in the modelling of Be stars \citep[e.g.][]{rimu2018, mota2019, suff2020, mota2021}. In this Section, we use these methods to determine which stellar parameters best reproduce observations from the IUE, previously described in Section \ref{sec:observations}. The UV spectrum is not strongly affected by the disk, so we used it for our fitting procedure to find the stellar parameters. We determine values for the stellar mass $M$, critical fraction of rotation $W$ \citep[as defined in equation 6 of][]{rivi2013}, age $t/t_{ms}$ (where $t_{ms}$ is the main sequence lifetime), inclination $i$, distance $d$ and the degree of interstellar reddening E($B-V$), and from these compute the derived parameters listed in Table \ref{tab:66ophpara}.

We use a grid of $770$ diskless Be star models, called BeAtlas, to create a parameter space for evaluating the observed spectrum from IUE. The BeAtlas grid was originally computed by \citet{mota2019} using a 3D non-local thermodynamic equilibrium (non-LTE) Monte Carlo radiative transfer code called \hdust\ \citep{carc2006}. This code simulates Be stars with or without disks by solving the coupled problems of radiative equilibrium, statistical equilibrium, and radiative transfer to compute synthetic observables. Table~\ref{tab:beatlasparams} summarizes the ranges of the stellar parameters for the BeAtlas models; the masses are consistent with those computed by \citet{geor2013}.

The parameter space was sampled using \emcee\ \citep{fore2013}, a Python code of the Affine Invariant MCMC Ensemble sampler \citep{good2010}. We defined Gaussian prior distributions with mean and variance taken from literature values (see Table \ref{tab:emceeprior}). We used the parallax from Hipparcos \citep{vanl2007} since the parallaxes from GAIA's DR2 catalogue \citep{gaiaDR2} contained large errors for bright stars, including 66 Oph. Recently, we found recomputing the stellar parameters using the parallax reported in GAIA’s eDR3 catalogue \citep{gaiaeDR3} ($\sim4.90\pm{0.37}~\rm{mas}$, or $\sim204\substack{+17 \\ -14}~\rm{pc}$) produced a consistent set of parameters.

Before fitting, each model was also artificially reddened using the standard \citet{fitz1999} parameterization, with E($B-V$) as a free parameter in determining each model's goodness of fit. We used a $\rm{log}(\chi^2)$ likelihood function to evaluate the fit of the models to the observations.

\begin{table}
	\centering
	\begin{tabular}{@{} l
                    @{\hspace*{4mm}}l
                    @{\hspace*{10mm}}l
                    @{\hspace*{4mm}}l
                    @{\hspace*{4mm}}l @{}}
		\hline
		\hline
		Best Fit &  & & Derived &  \\
		Parameters & Values & & Parameters & Values \\
		\hline
		$M$ [$\rm{M_{\odot}}$]  &  $11.03\substack{+0.55 \\ -0.53}$ & & $L$ [$\rm{L_{\odot}}$]  &  $8200\substack{+1600 \\ -1300}$ \\
		$W$  &  $0.52\substack{+0.05 \\ -0.05}$ & & $T_{eff}$ [K]  &  $25940\substack{+800 \\ -750}$ \\
		$t/t_{ms}$  &  $0.33\substack{+0.10 \\ -0.08}$ & & $\log g$  &  $4.17\substack{+0.05 \\ -0.06}$ \\
		$i$ [$\rm{^{\circ}}$]  &  $57.6\substack{+6.8 \\ -6.8}$ & & $R_{pole}$ [$\rm{R_{\odot}}$] & $4.50\substack{+0.32 \\ -0.23}$ \\
		$d$ [pc]  &  $208.6\substack{+8.9 \\ -9.0}$ & & $R_{eq}$ [$\rm{R_{\odot}}$] & $5.11\substack{+0.37 \\ -0.26}$ \\
		E($B-V$)  &  $0.22\substack{+0.01 \\ -0.01}$ & & $v {\sin} (i)$ & $290\substack{+11 \\ -9}$ \\
		\hline
		\hline
	\end{tabular}
	\caption{The best fitting stellar parameters for 66 Oph computed with \emcee, and derived from the computed values using the models of \citet{geor2013}.}
	\label{tab:66ophpara}
\end{table}

\begin{table}
	\centering
	\begin{tabular}{lllll}
		\hline
		\hline
		Parameter & Grid Values & \\
		\hline
		$M$ [M$_{\odot}$]  &  $1.7, 2, 2.5, 3, 4, 5, 7, 9, 12, 15, 20$  &  \\ 
		$W$         &  $0.00, 0.33, 0.47, 0.57, 0.66, 0.74, 0.81, 0.93, 0.99$  &  \\ 
		$t/t_{ms}$  &  $0, 0.25, 0.5, 0.75, 1, 1.01, 1.02$  &  \\  
		\hline
		\hline
	\end{tabular}
	\caption{The mass, critical fraction of rotation and age used for the BeAtlas grid of models.}
	\label{tab:beatlasparams}
\end{table}

\begin{table}
	\centering
	\begin{tabular}{lllll}
		\hline
		\hline
		Parameter & Value & Reference & \\
		\hline
		parallax [mas]  &  $5.01 \pm 0.26$  & \citep{vanl2007} &  \\  
		$v$sin$(i)$ [km/s]  &  $280 \pm 17$  & \citep{gran2010} &  \\  
		$i$ [\textsuperscript{$\circ$}] &  $43 \pm 8$ & \citep{floq2002} &  \\ 
		\hline
		\hline
	\end{tabular}
	\caption{The values used as prior distributions in \emcee\ .}
	\label{tab:emceeprior}
\end{table}

\begin{figure*}
	\centering
	\makebox[\textwidth][c]{\includegraphics[width = 1\textwidth]{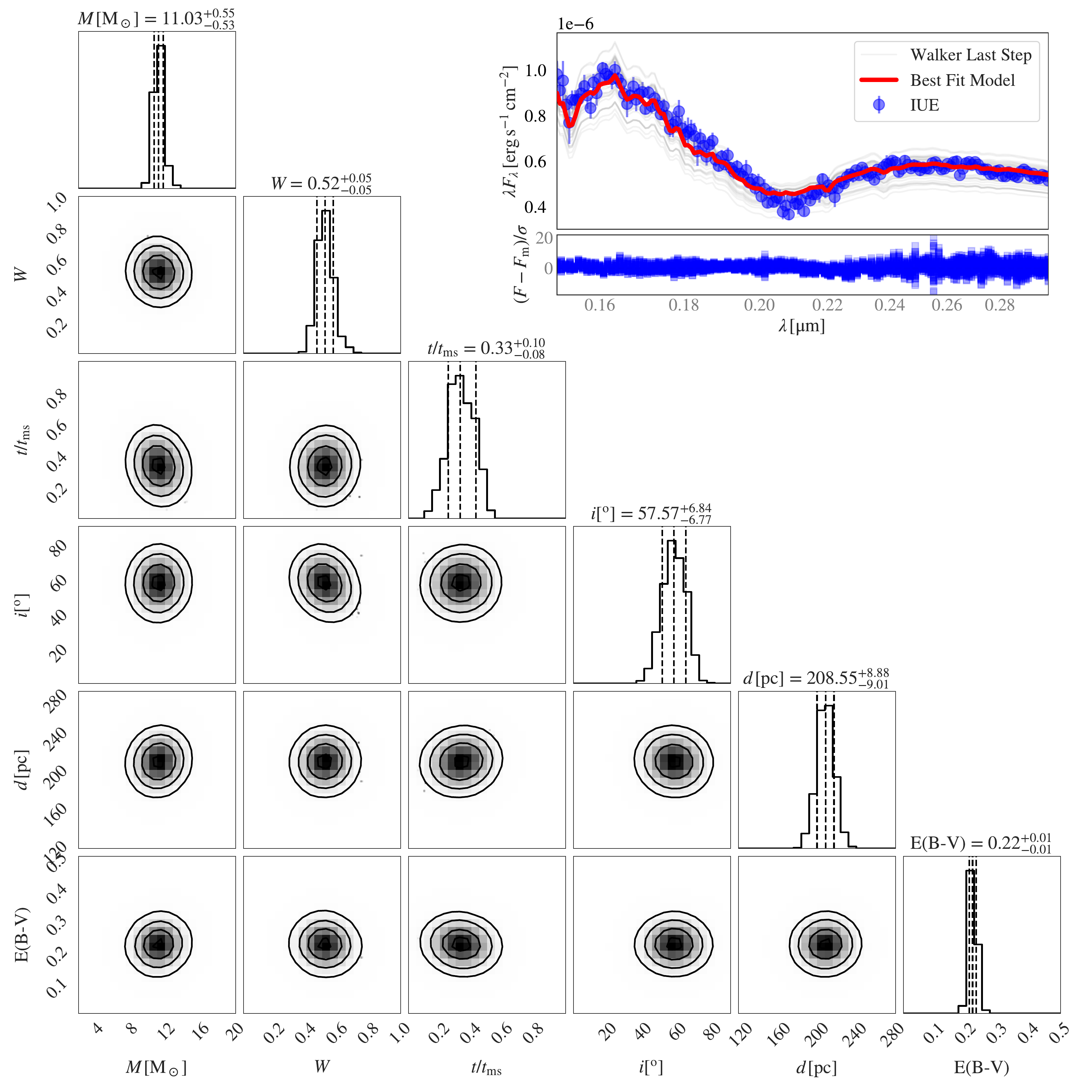}}
	\caption{The best fitting stellar parameters to 66 Oph's UV spectrum. The probability density functions of each parameter are shown on the main diagonal axis while the intersection for each parameter shows the correlation map. The six parameters included in the fitting procedure are the stellar mass $M$, the critical fraction of rotation $W$, time of life on the main sequence $t/t_{ms}$, stellar inclination $i$, distance $d$, and interstellar reddening E($B-V$). The subfigure in the top-right corner shows the UV data, the model corresponding to the best fit parameters and the predictions from the last step of each walker. The residuals between the data and model are shown directly below this subfigure.}
	\label{fig:bemcee}
\end{figure*}

Our \emcee\ computation used $30$ walkers and $50000$ steps with a burn-in of $5000$ steps, which were chosen by following the guidelines discussed by \citet{fore2013}. Approximately $27\%$ of the steps proposed by the MCMC were accepted, which is consistent with the range recommended by \citet{fore2013}, when using a walker step size of $3.0$.

Figure~\ref{fig:bemcee} shows the resulting probability density functions (PDFs) for the stellar parameters and correlation between these parameters as predicted by \emcee. The top right corner of Figure \ref{fig:bemcee} shows UV data along with a model computed with the best fit parameters in red. The grey lines show the model predictions corresponding to the last step of each walker. The residuals between the data and the model are shown directly below this panel.

The stellar parameters predicted by \emcee\ are used to compute the mean stellar luminosity $L$, mean effective temperature $T_{eff}$, mean surface gravity $g$, polar radius $R_{pole}$, equatorial radius $R_{eq}$, and $v {\sin} (i)$ by interpolating the Geneva stellar evolutionary models from \citet{geor2013} \citep[more details are available in][]{mota2019}. Table~\ref{tab:66ophpara} lists the most probable stellar parameters and the additional parameters derived from the Geneva models. The mass, temperature and radius are consistent with standard values for B0 to B2 stars \citep{cox2000}, and are in agreement with \citet{wate1987}, \citet{floq2002}, \citet{frem2006} and \citet{viei2017}. Values of $v {\sin} (i)$ reported in the literature range from $200~\rm{km/s}$ \citep{bern1970} to $292~\rm{km/s}$ \citep{floq2002}; our derived value agrees more closely with the upper limit. Our stellar parameters are further confirmed in Subsection \ref{subsec:results_1}, by comparison with the observed diskless SED.

\section{Disk Structure}\label{sec:results}

In Subsection~\ref{subsec:results_1} we determine the quasi-steady state of the disk by SED fitting and H$\alpha$ line modelling prior to the dissipation. We model the H$\alpha$ observation obtained by the HPOL Be star campaign on June 28, 1989. This observation \citep[previously discussed by][]{drap2014} corresponds to our determination of the onset of disk dissipation in March 1989. This observation was obtained at a key time when 66 Oph's disk was neither growing nor dissipating. The evolution of the disk before and after the quasi-steady state is modelled in Subsection~\ref{subsec:results_2}. These models consider four different scenarios: whether the disk is isothermal or non-isothermal, and whether the disk builds to a steady state then dissipates, or dissipates from the steady state. These scenarios are presented as follows; Subsection~\ref{subsubsec:results_2_1} -- isothermal disks that build then dissipate, Subsection~\ref{subsubsec:results_2_2} -- isothermal disks that dissipate only, Subsection~\ref{subsubsec:results_2_3} -- non-isothermal disks that build then dissipate, and Subsection~\ref{subsubsec:results_2_4} -- non-isothermal disks that dissipate only. Models for each scenario were constrained by SED and H$\alpha$ line fitting, and further compared to polarization when appropriate.

\subsection{Modelling the Quasi-Steady State Phase}\label{subsec:results_1}

In the VDD model, gas is injected into the inner disk and diffuses outward via viscous forces \citep{port99}. VDDs were initially modelled by \citet{port99} and later by \citet{okaz2001}, \citet{bjork2005}, \citet{jones2008} and \citet{lee13}. These studies showed that with a constant mass injection rate the disk's density profile becomes relatively constant as a steady state is reached over months, years or longer, since gas that is moving outward is replenished at the innermost boundary. However, \citet{haub2012} claims that a true steady state is never realized but that a quasi-steady state phase is observed in many Be stars \citep[e.g.][]{klem15}.

The H$\alpha$ EW did not vary significantly between 1985 and 1995 (as seen in the middle panel in Figure \ref{fig:allplots}). During this time, 66 Oph's disk was approximately quasi-steady, however the drop in V-band magnitude and polarization, which form closer to the star \citep{faes2013}, suggest that the dissipation began between 1989 and 1990 (indicated by the vertical, solid grey line in Figure \ref{fig:allplots}).

Our initial goal is to quantify the density structure of 66 Oph's disk while in this quasi-steady state so that we can use this model as a baseline to follow subsequent dissipation.

\begin{figure}
	\centering
	\makebox[\columnwidth][c]{\includegraphics[width = \columnwidth]{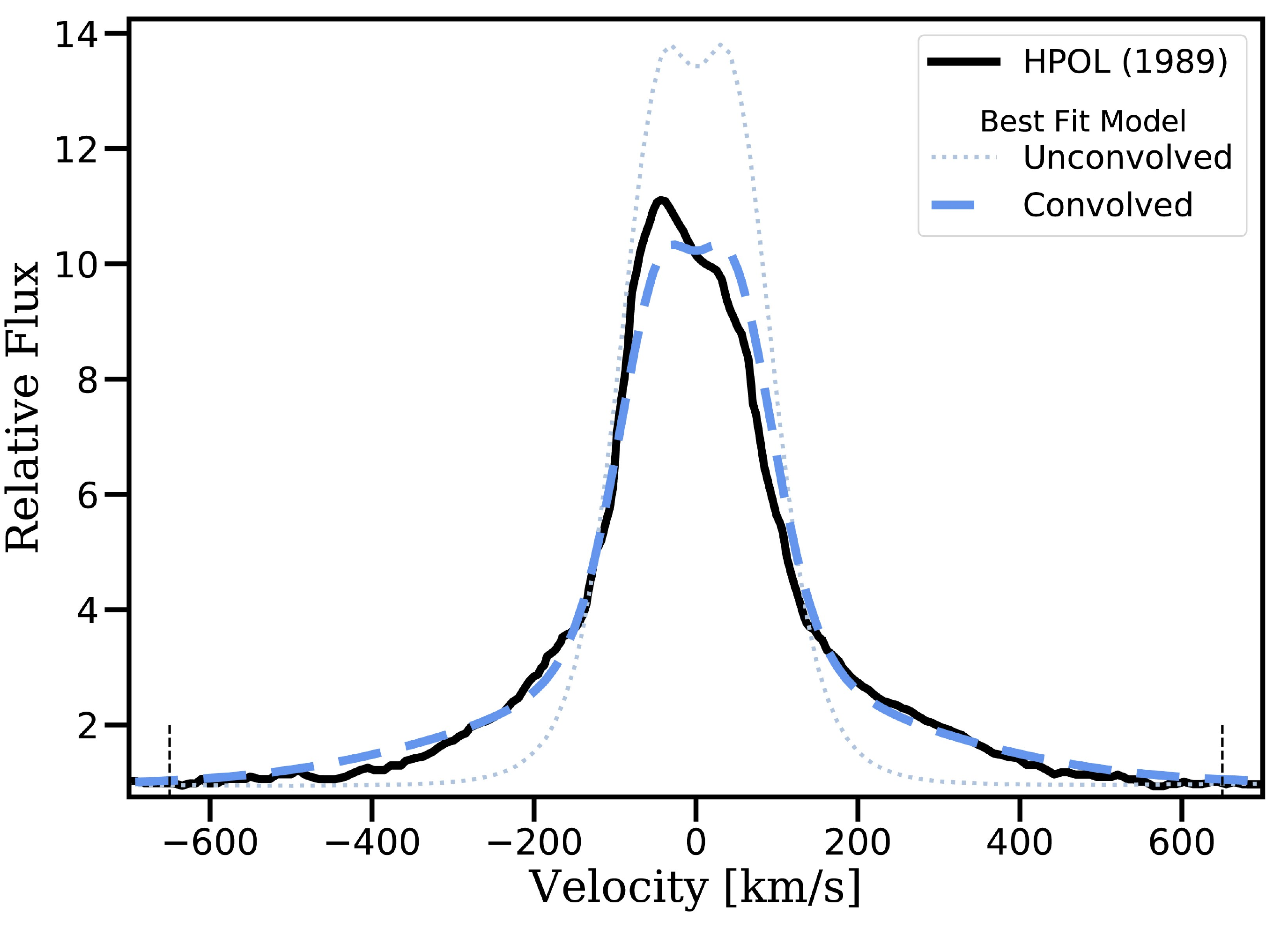}}
	\caption{A comparison of the HPOL 1989 H$\alpha$ profile to the best fit unconvolved model (dotted, grey) and final convolved spectra (dashed, blue). The vertical dashed lines show the range over which the fit was evaluated.}
	\label{fig:mcline_halpha_fit}
\end{figure}

Often in the literature, a power-law density distribution is used to represent a snapshot of the disk in time. We use the density structure originally adopted by \citet{wate1986} for IR continuum observations,
\begin{equation}
    \rho(r) = \rho_0\bigg(\frac{R_{eq}}{r}\bigg)^{n}.
    \label{eqn:rho_n}
\end{equation}
Here, $\rho_0$ is the density at the innermost disk radius in the equatorial plane, $r$ is the radial distance from the central star, and $n$ is the power law exponent.

We computed a grid of $61568$ models with densities from $\rho_0 = 10^{-10}$ to $10^{-13}~\rm{g\ cm}^{-3}$ for every quarter magnitude, with $n = 2.0$ to $3.5$ in steps of $0.1$, inclination from $i = 0$ to $90^{\circ}$ in steps of $2.5^{\circ}$, and outer disk radii of $R_{out} = 20$, $30$, $40$, $50$, $60$, $80$, $100$ and $200~\rm{R_{eq}}$. For each model, \hdust\ (previously described in Section \ref{sec:stellarparams}) was used to compute the SED and H$\alpha$ line profile.

\begin{figure}
	\makebox[\columnwidth][c]{\includegraphics[width = \columnwidth]{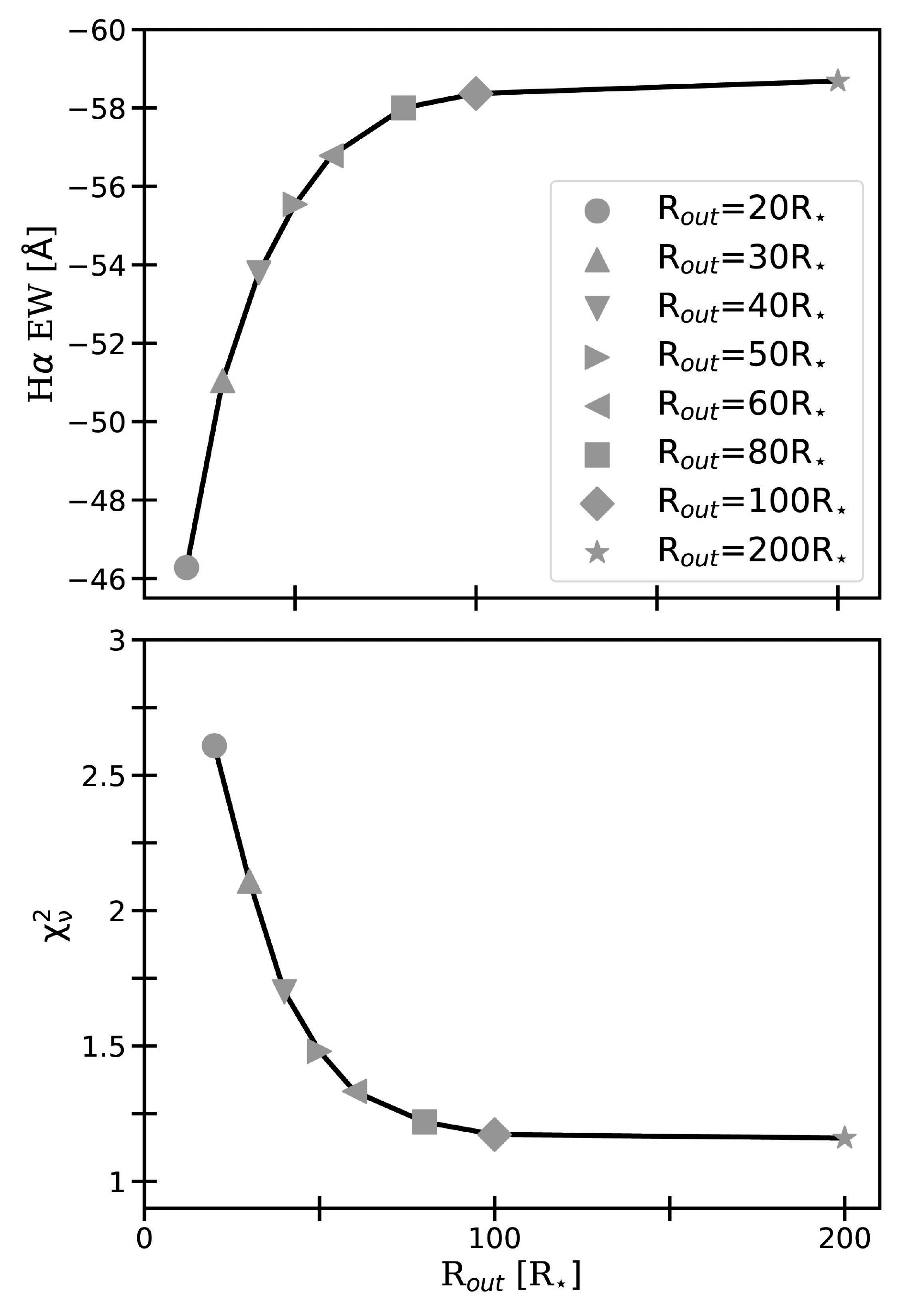}}
	\caption{Change in the H$\alpha$ EW and corresponding $\chi_\nu^2$ values for the best fit model with increases in outer disk radius. $\chi_\nu^2$ was computed  as the fit of the convolved profile to the HPOL 1989 line profile.}
	\label{fig:chi2_rout}
\end{figure}

Non-coherent electron scattering within the disk also affects the shape of line profiles by broadening the wings of the line \citep{humm1992}. This process is not accounted for in \hdust, so to approximate the effect, we assume that a fraction $f$ of the H$\alpha$ flux is scattered by electrons with thermal velocities ranging from $v_e = 300$ to $800~\rm{km/s}$, and that the scattered flux has a Gaussian profile. Thus, a new line profile $F_{\rm new}$ is obtained by
\begin{equation}
F_{\rm new}(\lambda) = (1-f) F_{\rm nc}\lambda + f \times F_{\rm c}(\lambda)\,,
\end{equation}
where $F_{\rm nc}$ is the non-convolved line profile predicted by \hdust\ and $F_{\rm c}$ is the convolution of $F_{\rm nc}$ with a Gaussian with FWHM of $v_e$. The convolved line profiles were then fit to the HPOL 1989 spectrum using \emcee\ (unique from the routine which used \emcee\ previously described in Section \ref{sec:stellarparams}). The fits were evaluated by interpolating the model fluxes to the wavelengths of the observed data and then calculating $\chi_\nu^2$ values.

\begin{figure*}[ht!]
	\centering
	\makebox[\textwidth][c]{\includegraphics[width = \textwidth]{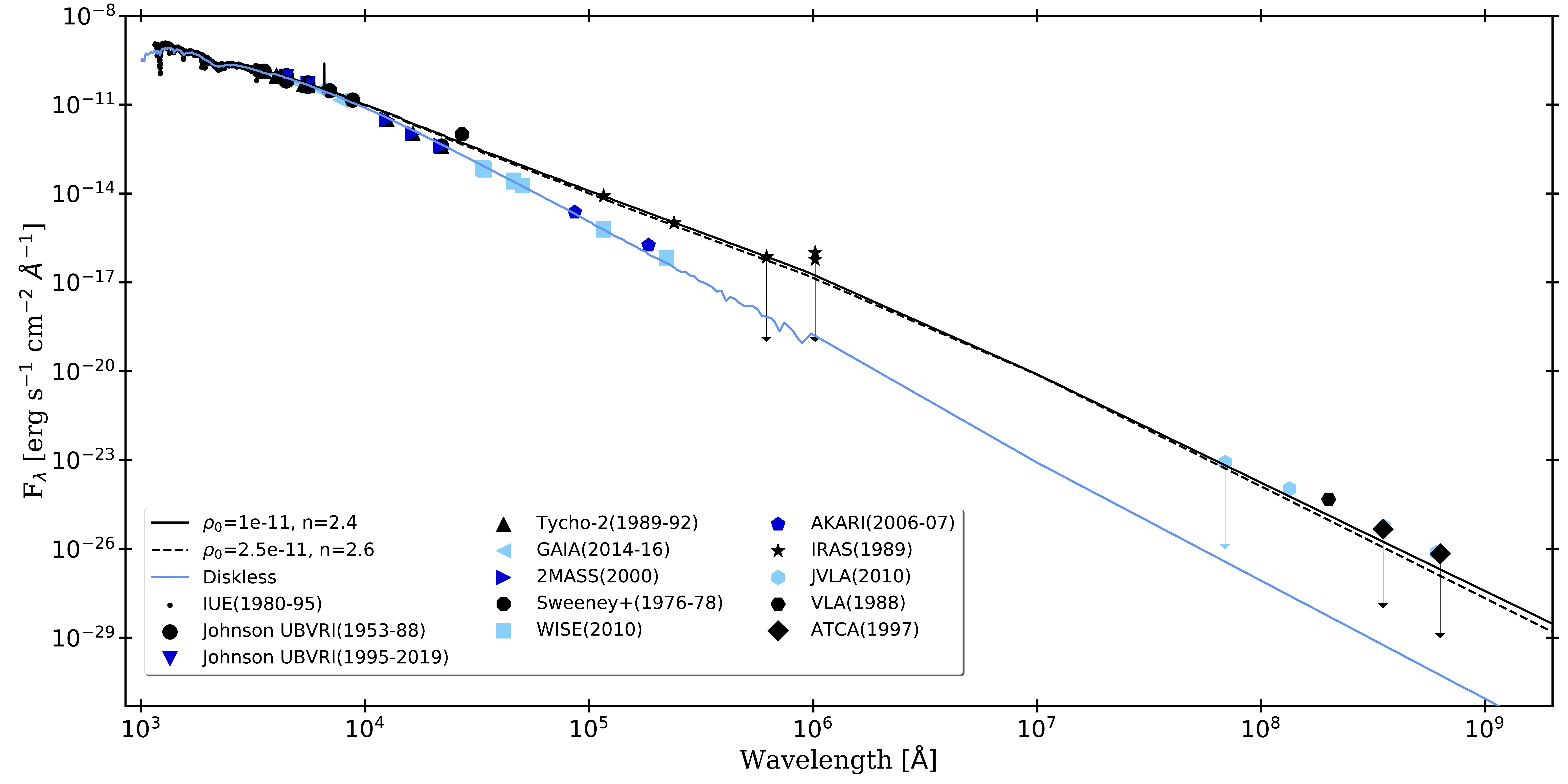}}
	\caption{A multi-epoch SED from UV through radio wavelengths, with the diskless and two best fitting disk simulations. Observation programs are differentiated by shapes. Colours differentiate evolutionary epochs; disk building (black), and dissipation before (dark blue) and after (light blue) H$\alpha$ went into absorption. Observations with downward arrows are considered upper limits \citep{wate1987, klem19}.}
	\label{fig:sed_merged}
\end{figure*}

After convolving, we found a subset of models within $1\sigma$ of the observed H$\alpha$ EW ($-58~\rm{\AA}$). The best fit model according to $\chi_\nu^2$ testing has a density of $\rho_0 = 2.5\times10^{-11}~{\rm g~cm}^{-3}$, with $n = 2.6$, at $i = 57.5^{\circ}$, and $R_{out} = 100~\rm{R_{eq}}$, with a fit of $\chi_{\nu}^2 = 1.17$ when using $F_c = 0.22$, $v_e = 700~{\rm km~s^{-1}}$. Figure~\ref{fig:mcline_halpha_fit} shows the observed profile (black line), and the best fit model before (dotted line) and after (dashed line) convolving. These profiles all have EW's of $-58~\rm{\AA}$. The thermal velocity used for the convolution corresponds to a kinetic temperature of $T_{K} = 10100~\rm{K}$. This agrees with the disk temperature of $\sim10500~\rm{K}$ at the edge of the optically thick H$\alpha$ emitting region at $\sim10.1~\rm{R_{eq}}$ \citep[computed by following Appendix D of][]{viei2017}. This process of convolving lines was previously used by \citet{klem15}, without MCMC fitting, for the Be star $\beta$ CMi.  $\beta$ CMi is a late-type star with a spectral type of B8Ve and is expected to have a cooler disk. We note that these authors obtained values of $f = 0.6$ and a smaller value of $v_e = 300~{\rm km~s^{-1}}$, consistent with its disk temperature.

In the case of a disk with $n=3.5$, the H$\alpha$ emitting volume probes a significant portion of the disk \citep{tycn2005} typically thought to be the innermost $20~\rm{R_{eq}}$ \citep{faes2013}. However, we found these regions become increasingly extended as $n$ decreases. This means that for an early-type star like 66 Oph, lower values of $n$ result in more ionized material at larger distances. Figure~\ref{fig:chi2_rout} shows the H$\alpha$ EW and corresponding $\chi_\nu^2$ value of the best fit model with increasing $R_{out}$. Here, with $n=2.6$, we find the EW increases until the disk reaches a radius of $\sim100~\rm{R_{eq}}$, beyond which there is no significant increase. Similarly, $\chi_\nu^2$ improves as $R_{out}$ increases and the H$\alpha$ flux increases until $\sim100~\rm{R_{eq}}$, with no appreciable change at larger radii. This provides an upper limit on the size of the H$\alpha$ emitting region.

The star's SED (Figure \ref{fig:sed_merged}) was also used to independently constrain our model values of $\rho_0$ and $n$. The SED upper bound, prior to dissipation, is set by IRAS (1989) in the IR, as well as radio observations from VLA (1988). Conveniently, the upper bound from IRAS was observed just before the onset of dissipation, providing important observational constraints for the quasi-steady state phase. Data from ATCA (1997) and JVLA (2010) are also on the upper bound, as the disk dissipation had not significantly changed the disk's radio emission which comes from the outermost disk. The SED's lower bound is constrained by observations from WISE (2010) and AKARI (2006-07).  Observations of the lower bound were obtained after dissipation and likely correspond to a diskless state. Using the diskless model of Section 3, we find a very reasonable fit to the observations at visible to radio wavelengths observed after 1989 with a $\chi^2_{\nu} = 1.12$.

\begin{figure}
	\centering
	\makebox[\columnwidth][c]{\includegraphics[width = \columnwidth]{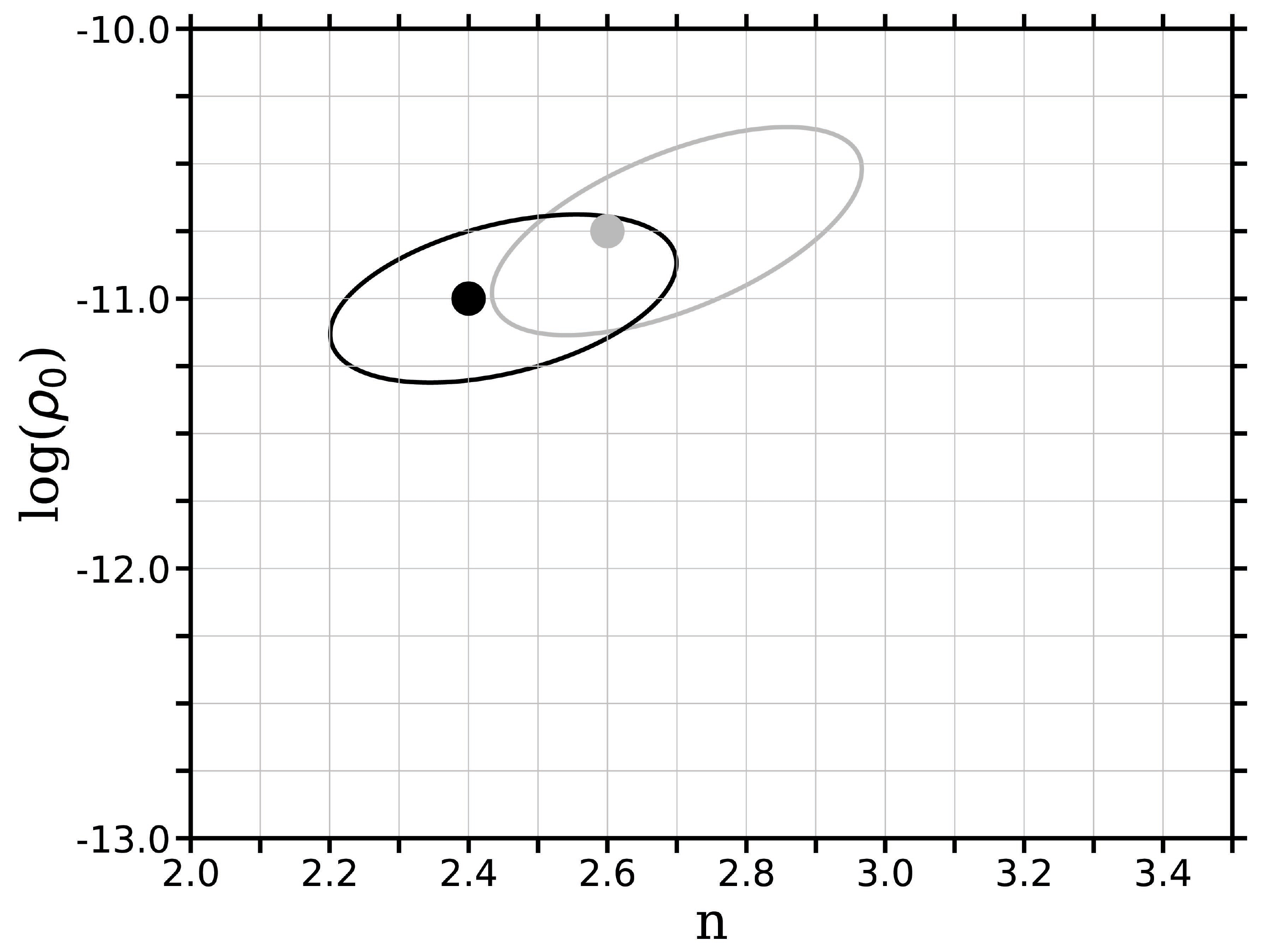}}
	\caption{Summary of the best fitting disk densities. Contours corresponding to the 1$\sigma$ confidence level were determined by interpolation of the $\chi^2_{\nu}$ values over the model grid. The black contours shows the SED fitting and the grey contour shows the H$\alpha$ fitting, with corresponding coloured dots indicating the best fit model.}
	\label{fig:chi2_rho_n}
\end{figure}

We fit the upper bound SED with the same grid of models used above for H$\alpha$ fitting. We find that a model of $\rho_0 = 10^{-11}~\rm{g~cm}^{-3}$, ${n = 2.4}$, $i = 57.5^{\circ}$ and $R_{out} = 100~\rm{R_{eq}}$ (shown as the solid black line in Figure~\ref{fig:sed_merged}, and the black dot in Figure~\ref{fig:chi2_rho_n}) produced the best fit to the visible, IR and radio photometry observed before 1989 with $\chi^2_{\nu} = 1.09$. Increasing the outer disk radius to $200~\rm{R_{eq}}$ marginally improved the fit to $\chi_{\nu}^2 = 1.07$. The model which best fit the H$\alpha$ line profile ($\rho_0 = 2.5\times10^{-11}~\rm{g~cm}^{-3}$ and ${n = 2.6}$) fit the SED observations before 1989 with $\chi^2_{\nu} = 1.15$ for $R_{out} = 100~\rm{R_{eq}}$, or $\chi^2_{\nu} = 1.14$ for $R_{out} = 200~\rm{R_{eq}}$.

Figure~\ref{fig:chi2_rho_n} shows a comparison of the best fit models to the H$\alpha$ line profile and SED according to the $\chi^2_{\nu}$ fitting. The contours show the models which fit best to within 1$\sigma$. The overlap has a range of $\rho_0 = 8.5\times10^{-12}$ to $3\times10^{-11}~{\rm g~cm}^{-3}$, and $n = 2.4$ to $2.7$. Within the same overlapping region, only the $\rho_0 = 2.5\times10^{-11}$~${\rm g~cm}^{-3}$ and $n = 2.6$ model satisfies both the SED and H$\alpha$ line profile fitting. The model which best fit the SED produced a relatively poor fit to the H$\alpha$ line profile, with $\chi^2_{\nu} = 3.2$.

The $\rho_0 = 2.5\times10^{-11}$~${\rm g~cm}^{-3}$ and $n = 2.6$ model predicts a V-band magnitude of $4.72~\rm{mag}$, which is consistent with the range of $4.76$ to $4.55~{\rm{mag}}$ observed in 1989 \citep{floq2002}. The same best model predicts the V-band polarization to be $\sim0.72\%$, while observations obtained 5 years prior to dissipation show the V-band polarization to range from $0.58\%$ to $0.65\%$ (after subtracting the interstellar polarization, as described in Section \ref{sec:observations}).

Since the $\rho_0 = 2.5\times10^{-11}$~${\rm g~cm}^{-3}$ and $n = 2.6$ model with $i = 57.5^{\circ}$ and $R_{out} = 100~\rm{R_{eq}}$ is clearly successful in reproducing the observed H$\alpha$ line profile, SED, V-band magnitude and V-band polarization, we conclude it is the best representation of the quasi-steady state phase for 66 Oph's disk.

\subsection{Hydrodynamic Calculations}\label{subsec:results_2}

In this section, we model the formation of 66 Oph's disk and the subsequent $21$ years of dissipation until the H$\alpha$ emission went into absorption. We use the best fit quasi-steady model from Section \ref{subsec:results_1} as a mid-point between the building and dissipation. To simulate the disk evolution, we use the 1D dynamical code \singlebe, developed by \citet{okaz2002} and later used by \citet{carc2012}, \citet{haub2012}, \citet{rimu2018} and \citet{ghor2018}.

\singlebe\ solves the 1D fluid hydrodynamic equations \citep{prin1981} for the surface density of a viscous disk using the thin disk approximation. The disk is assumed to be axisymmetric, Keplerian, and in vertical hydrostatic equilibrium. See \citet{okaz2002} and \citet{rimu2018} for more details on \singlebe.

The evolution of VDDs is controlled by kinematic viscous forces within the disk. \singlebe\ adopts the commonly used $\alpha$-prescription for viscosity, defined by \citet{shak1973} as

\begin{equation} 
    \nu(r) = \frac{2}{3}\alpha(r) c_{s}(r) H(r),
    \label{eqn:alphass}
\end{equation}

\noindent
where $\nu$ is the disk viscosity, $r$ is the radius in the disk, and $H$ is the disk scale height. The sound speed is given by

\begin{equation}
    c_s(r) = [(k_BT(r))/(\mu m_a)]^{1/2},
    \label{eqn:soundspeed}
\end{equation}

\noindent
where $k_B$ is the Boltzmann constant, $T$ is the gas kinetic temperature, $\mu$ is the mean molecular weight, and $m_a$ is the atomic mass unit. Note, for a non-isothermal disk, the radial dependence of the temperature can change the sound speed, which in turn changes the viscosity in Equation \ref{eqn:alphass}. In this case, $\alpha(r)$ and $T(r)$ become degenerate and indistinguishable. If one parameter were to be held constant, the other parameter may be adjusted to produce the same viscosity. Therefore, we distinguish isothermal disks from non-isothermal disks based on constant or variable $\alpha T$ with radius, respectively.

Viscosity affects disk evolution by influencing the outflow rate of the gas \citep{lee91}. Given $\alpha(r)$, a target inner surface density $\Sigma(r=R_{eq})$ (i.e. $\Sigma_0$) or inner volume density $\rho_0$, the outer disk radius $R_{out}$, and a value for the mass injection rate at the base of the disk $\dot{M}_{inj}$, \singlebe\ computes the surface density as a function of time and distance from the star, $\Sigma(r,t)$. Details about the computational procedure and assumed boundary conditions can be found in \citet{rimu2018}.

Using equation 3.6 of \citet{okaz1991}, the mass density can be computed as

\begin{equation}
    \rho(r,t) = \frac{\Sigma(r,t)}{\sqrt{2\pi}H(r,t)}.
\end{equation}

\noindent
We note that, assuming a constant $\alpha T$, hydrodynamic theory for an axisymmetric disk with Keplerian rotation in a quasi-steady state predicts that the mass density will have approximately a radial power law (Equation \ref{eqn:rho_n}) with ${n = 3.5}$ \citep{bjork2005}.

We input the stellar parameters  determined in Section \ref{sec:stellarparams} (see Table \ref{tab:66ophpara}),  $\rho_0$, $n$, and $R_{out}$ of the best fit model from Subsection \ref{subsec:results_1}, and the duration of the growth and dissipation events to \singlebe.  We use $\alpha = 1.0$ to ensure rapid evolution \citep{haub2012}, however as $\alpha$ can be scaled by changing the length of the evolutionary epoch \citep[as illustrated in figure 1 of][]{haub2012} our models can explore all values of $\alpha$. For each model, the disk's innermost radius was set to $1~\rm{R_{eq}}$ and the radial grid points were spaced logarithmically from the inner radius. Following \citet{rimu2018}, we define the onset of dissipation to occur at $0$ years for all models. The density profiles from \singlebe\ were input to \hdust\ to compute the SED, H$\alpha$ line profile, and polarization at times corresponding with the observed data.

\subsubsection{Building and Dissipating an Isothermal Disk}\label{subsubsec:results_2_1}

\begin{figure}
	\makebox[\columnwidth][c]{\includegraphics[width = 1\columnwidth]{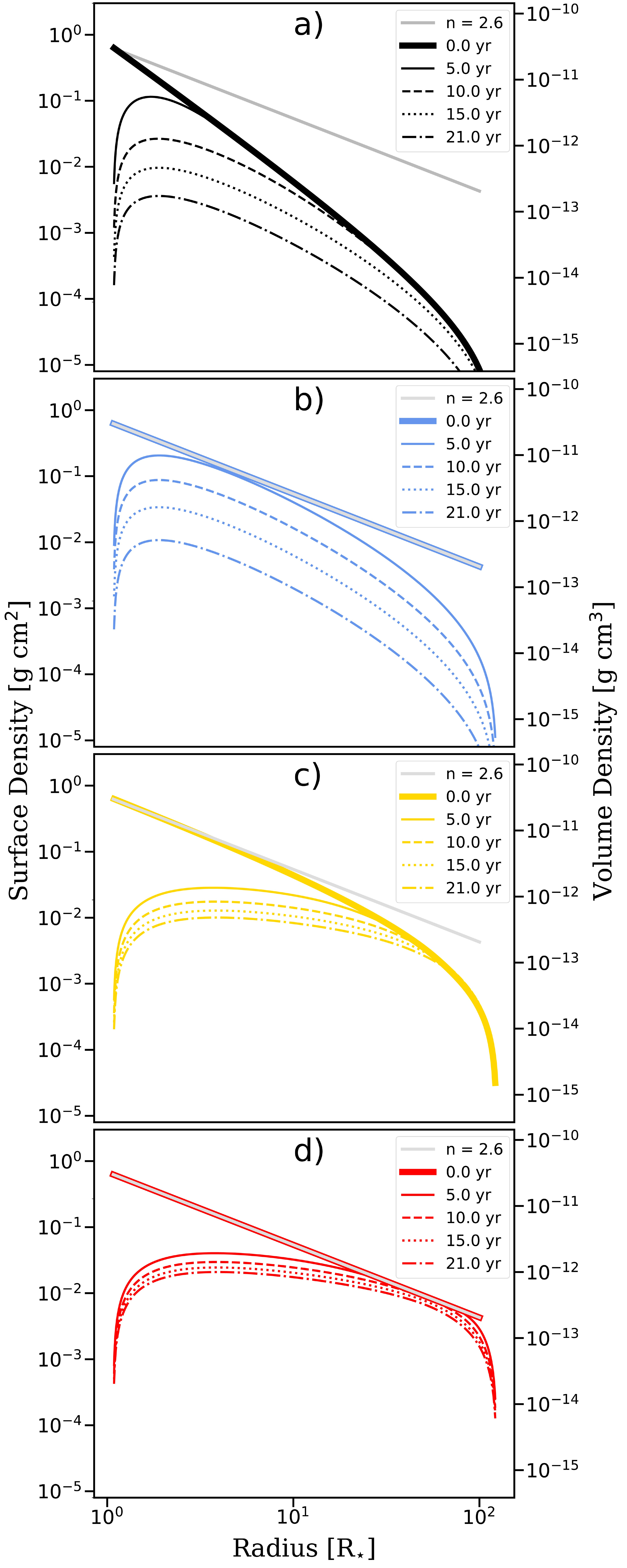}}
	\caption{Disk density profiles at different times from the start of dissipation for the following scenarios: Panel a) constant $\alpha T$, $n=3.5$, \textit{b)} constant $\alpha T$, $n=2.6$, \textit{c)} variable $\alpha T$, $n=2.7$, \textit{d)} variable $\alpha T$, $n=2.6$. A grey line in each panel indicates the best quasi-steady model.}
	\label{fig:sbe_triplot}
\end{figure}

\begin{figure*}[t!]
    \centering
	\makebox[\textwidth][c]{\includegraphics[width = 1\textwidth]{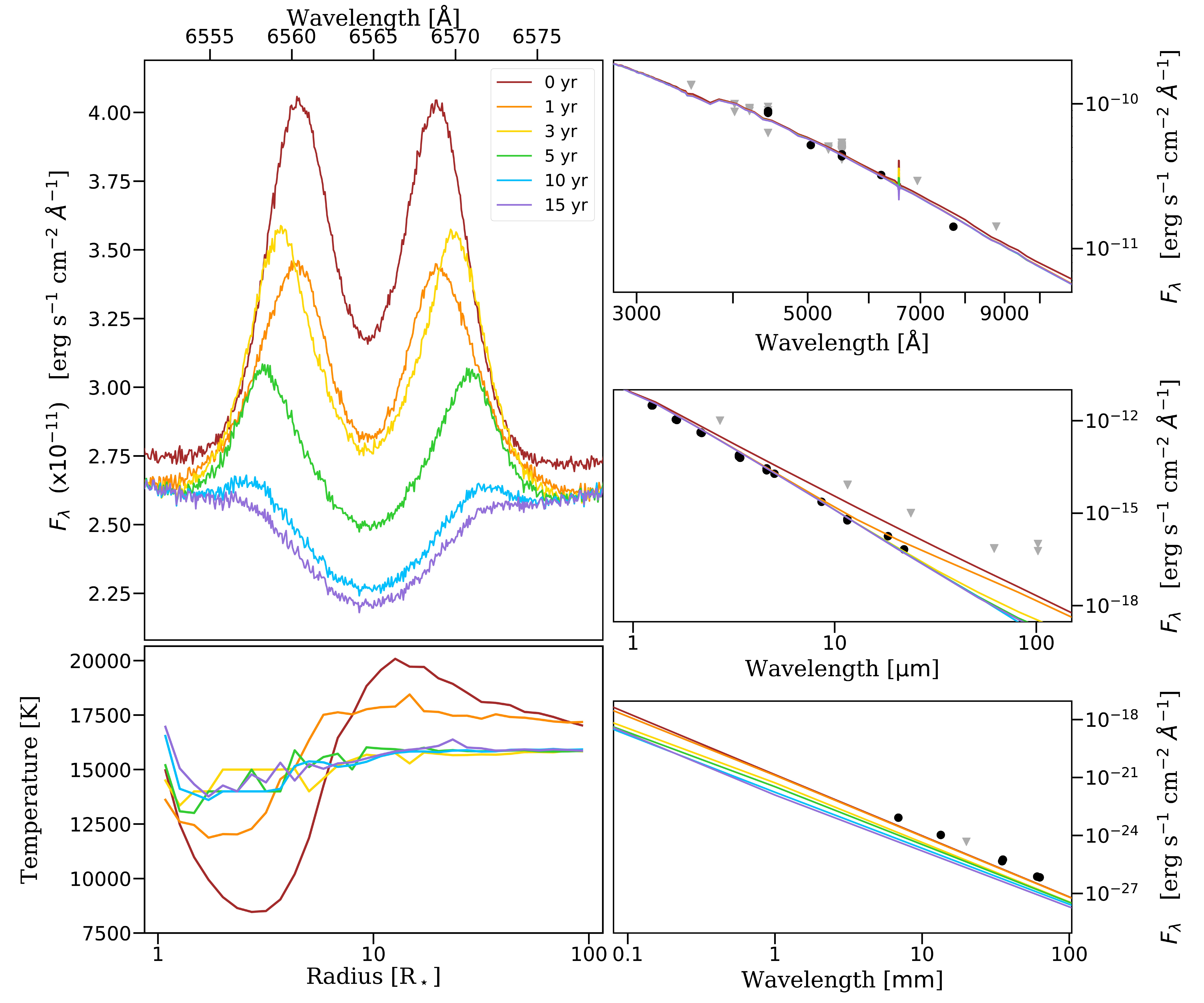}}
	\caption{Changes in the H$\alpha$ line (top left), the disk temperature structure (bottom left) and the visible (top right), IR (middle right) and radio (bottom right) SEDs during the disk dissipation of the constant $\alpha T$, $n = 3.5$ model. The H$\alpha$ spectra are not continuum normalized to clearly show the increase in peak flux during dissipation. The SEDs include observations made before 1989 (grey triangles) and after 1989 (black circles) from Figure \ref{fig:sed_merged}.}
	\label{fig:sbe_n35_a10}
\end{figure*}

Here, we investigate an isothermal disk which builds to a steady state and then dissipates, which we refer to as the constant $\alpha T$, $n=3.5$ scenario. The isothermal assumption also makes our models similar to the mixed models used in \citet{carc2008}, where the disk is isothermal in the fluid equations and variable in temperature when computing observables using \hdust.

To match the base density found in Section \ref{subsec:results_1}, this scenario requires a mass injection rate of $\dot{M} = 8.5\times10^{18}~\rm{g/s}$ (or $\sim1.3\times10^{-7}~\rm{M_{\sun}/yr}$) at $R = 1.02 R_{eq}$, of which $\sim99$\% falls back onto the star. We chose a constant disk temperature at $60\%$ of $T_{eff}$ (from Table \ref{tab:66ophpara}), following \citet{carc2006}.

Panel a) of Figure \ref{fig:sbe_triplot} shows density profiles of this scenario after the building period during the dissipation. The quasi-steady state is indicated by the thick, black line on the panel. We note that it has the expected value of $n = 3.5$  which differs from the ${n=2.6}$ of the best fitting quasi-steady model. During dissipation, the disk empties in an ``inside-out'' fashion, which is expected for Be stars as discussed by \citet{poeck1982}, \citet{rivi2001} and \citet{clark2003}. After the dissipation, the maximum density of the disk is $\sim5\times10^{-13}$~${\rm g~cm}^{-3}$.

Figure \ref{fig:sbe_n35_a10} shows the H$\alpha$ line profiles, the disk temperature profiles, and the visible, IR and radio SEDs compared to observed data (previously shown in Figure \ref{fig:sed_merged}), at different times during dissipation. The SEDs in the three right panels show that this scenario is unable to reproduce the observed flux before dissipation. This is most notable at IR wavelengths where the flux is $\sim1$ to $2$ orders of magnitude fainter. In the top left panel of Figure \ref{fig:sbe_n35_a10}, the H$\alpha$ line at $0~\rm{years}$ shows a peak flux of $\sim1.5$ times the continuum flux. In contrast, the HPOL 1989 observation (previously shown in Figure \ref{fig:mcline_halpha_fit}) shows a peak flux of $\sim11$ times the continuum flux.

The ${n = 3.5}$ disk also has a local maximum of the peak H$\alpha$ flux at $3~\rm{years}$ after the dissipation started. This is due to an increase of stellar ionizing radiation (and corresponding increase in H$\alpha$ flux) reaching larger radial distances of the inner disk due to the reaccretion of material closest to the star. (See the bottom left panel of Figure~\ref{fig:sbe_n35_a10}.) After this, the disk continues to dissipate and the H$\alpha$ flux drops smoothly. This phenomenon was also observed for disk models with $\alpha = 0.1$ and $\alpha = 0.5$, with the flux increase occurring at later times in the dissipation.

\begin{figure*}
    \centering
	\makebox[\textwidth][c]{\includegraphics[width = 1\textwidth]{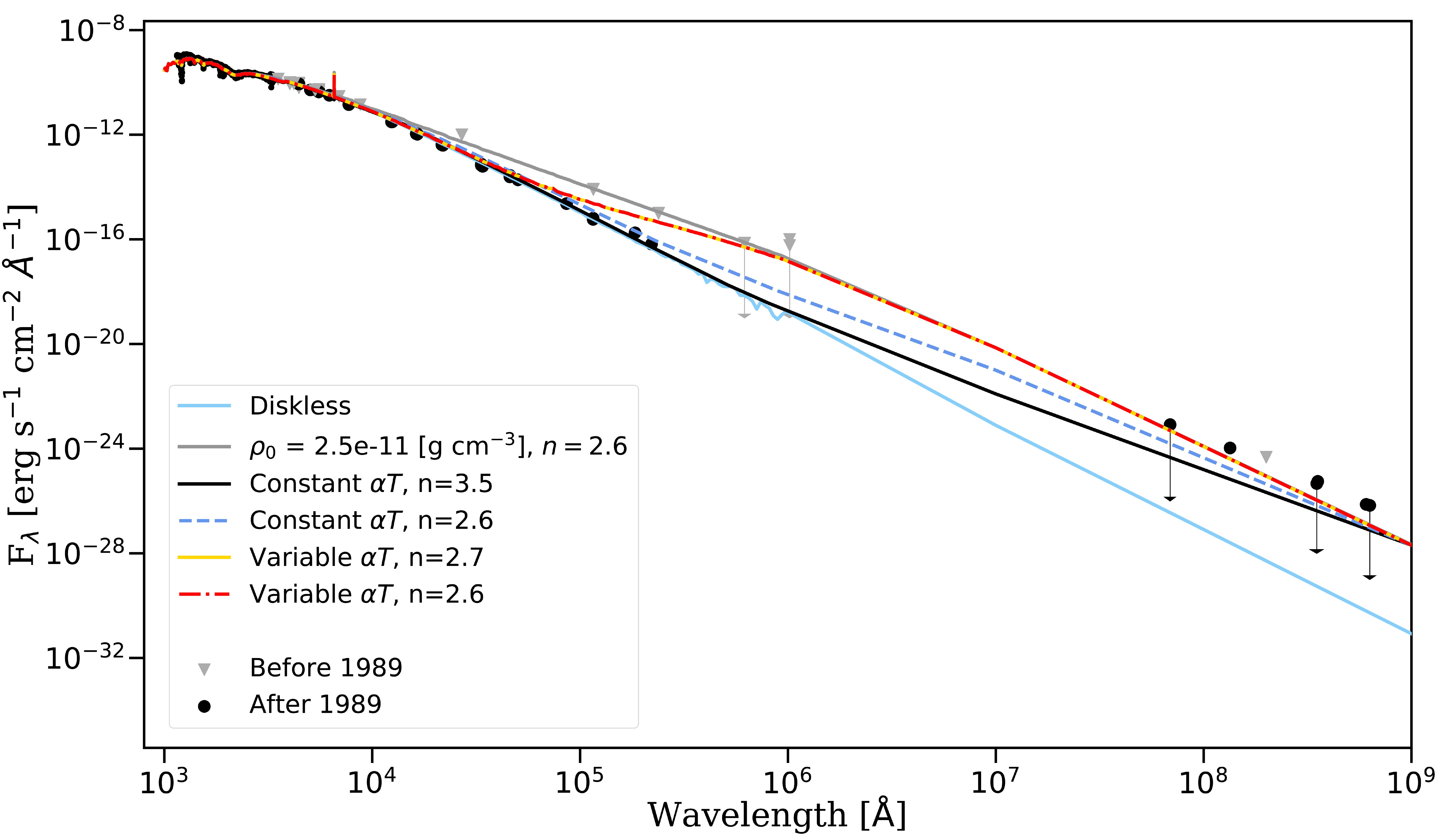}}
	\caption{SEDs for each of the dynamical scenarios after dissipation. For comparison the best fitting initial steady state is shown. The observed data is the same as that presented in Figure \ref{fig:sed_merged}. The two models from the variable $\alpha T$ scenarios lie on top of each other. The colours for each model are consistent with Figure \ref{fig:sbe_triplot}.}
	\label{fig:sbe_sed}
\end{figure*}

In Figure~\ref{fig:sbe_sed} we show the SED for this scenario after $21$ years of dissipation (black line), and also the best quasi-steady state model (grey line). As the disk dissipates, the model approaches the theoretical diskless SED. This model closely fits the UV and IR observations, however produces a weaker radio excess than the 2010 JVLA observations.

Figure~\ref{fig:sbe_n2.5_halpha} shows the H$\alpha$ line profile (black line) for this scenario after $21$ years of dissipation. The line profile shows no emission and is consistent with the absorption profile of the NRES/LCOGT 2020 observation. Note, the NRES/LCOGT observation closely matches with the observations from BESS obtained in 2010.

Figure~\ref{fig:ew_sim_obs} shows the H$\alpha$ EW during dissipation for this scenario (black line). Here it is apparent that the disk could only build to approximately half of the observed EW of the quasi-steady state and dissipates at a much faster rate than the observations.

In conclusion, the constant $\alpha T$, $n=3.5$ scenario fails to reproduce the two main observables chosen to constrain the models, namely the SED and H$\alpha$ line emission. This is further illustrated in Figure \ref{fig:sbe_triplot}, panel a), which shows that this model produces a disk far less massive than what is required. This will be further discussed below.

\subsubsection{Dissipating an Isothermal Disk from a Defined Density}\label{subsubsec:results_2_2}

Here we use an ad hoc scenario that starts to dissipate from the quasi-steady state with ${n = 2.6}$ in order to bypass the building phase. This constant $\alpha T$, $n = 2.6$ scenario, uses the same $\alpha$ and $T_{disk}$ as the constant $\alpha T$, $n = 3.5$ scenario.

Figure~\ref{fig:sbe_triplot}, panel b) shows that as dissipation begins, this scenario quickly moves toward a steeper density slope. During the first five years, the gas from the outer disk moves to the inner disk, preventing the density of the disk between $\sim5$ and $10~\rm{R_{eq}}$ from depleting. As dissipation continues, the entire disk appears to dissipate at a steady rate across all radii.

Figure \ref{fig:sbe_sed} shows the SED for this scenario (blue dashed line) closely matches the visible and IR observations after $21$ years of dissipation. An excess of radio flux is maintained as gas remains in the outer disk.

\begin{figure}
    \centering
	\makebox[\columnwidth][c]{\includegraphics[height=0.32\textheight,width = 1\columnwidth]{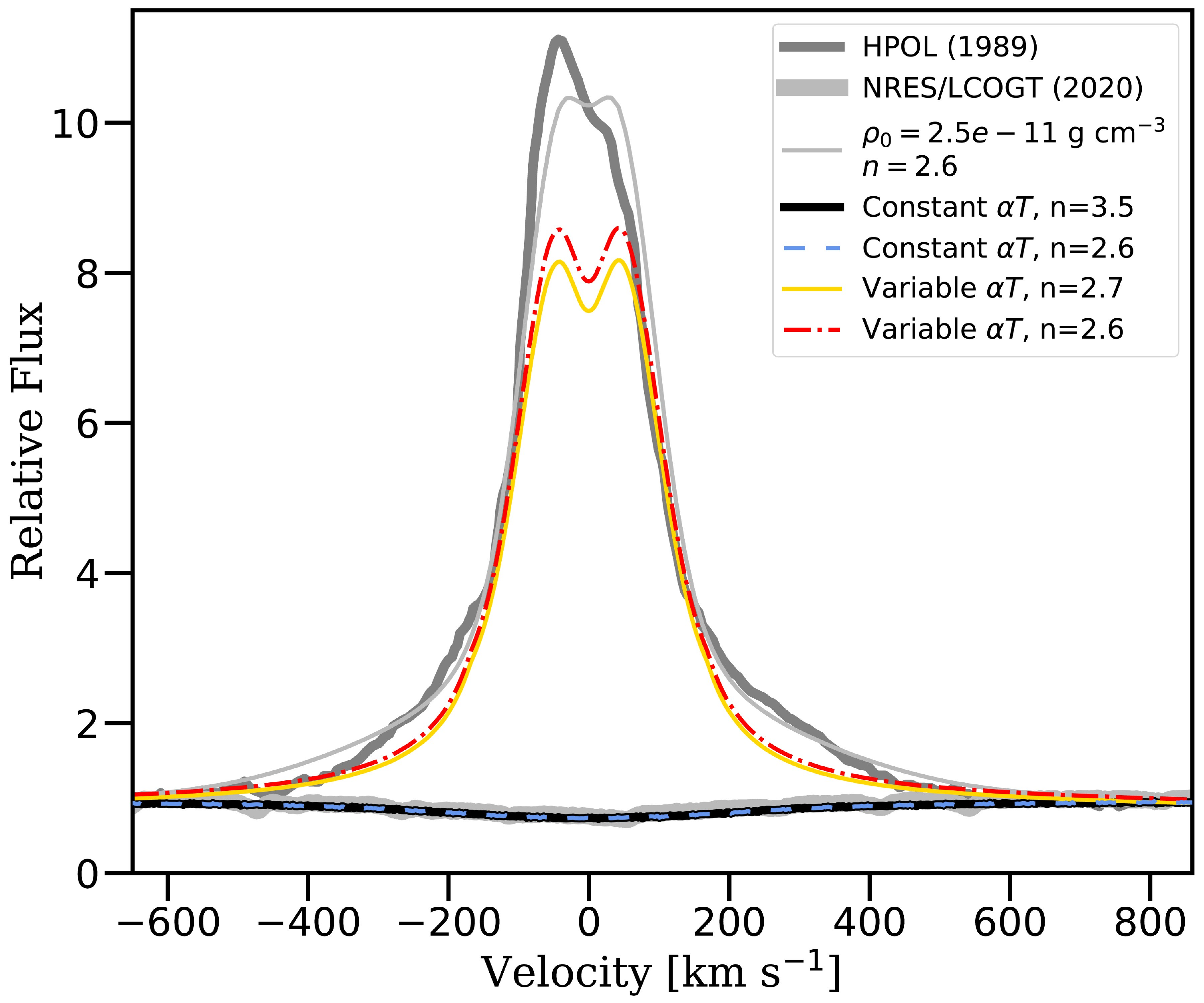}}
	\caption{The same as Figure \ref{fig:sbe_sed} for the H$\alpha$ line profile. The simulated lines were convolved using the Gaussian determined to give the best fit in Figure~\ref{fig:mcline_halpha_fit}. The models from the constant $\alpha T$, $n=3.5$ and $n=2.6$ scenarios lie on top of the NRES/LCOGT 2020 observation and in absorption.  The colours for each model are consistent with Figure \ref{fig:sbe_triplot}.}
	\label{fig:sbe_n2.5_halpha}
\end{figure}

\begin{figure}
    \centering
	\makebox[\columnwidth][c]{\includegraphics[width = 1\columnwidth]{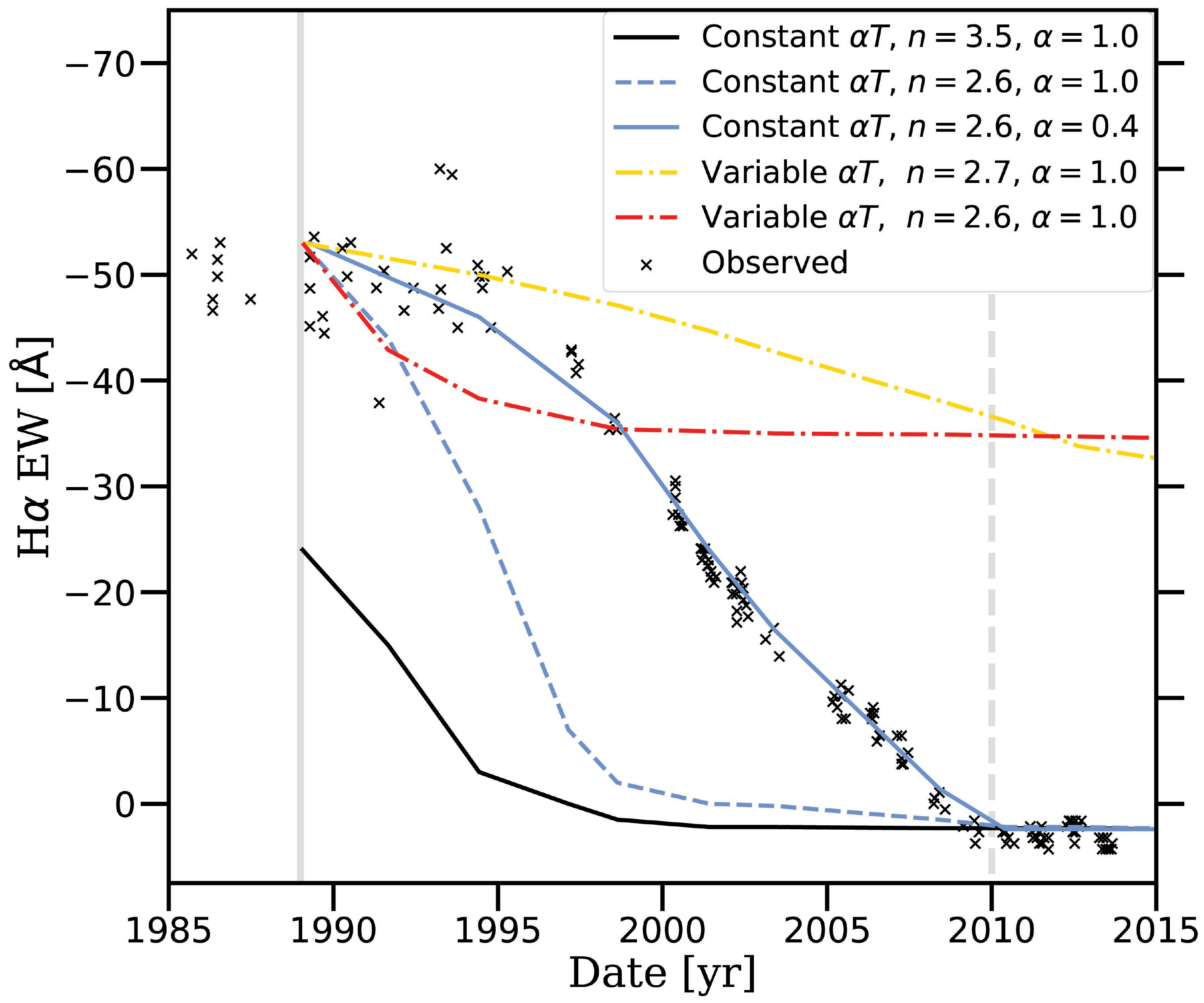}}
	\caption{The same as Figure \ref{fig:sbe_sed} for the H$\alpha$ EW. The solid vertical grey line indicates the onset of dissipation, and the dashed grey line indicates when the H$\alpha$ line was observed to transition to absorption. While the model parameters differ, the constant $\alpha T$, $n = 2.6$ scenario with $\alpha = 0.4$ and $R_{out} = 115~\rm{R_{eq}}$ is shown for comparison. The colours for each model are consistent with Figure \ref{fig:sbe_triplot}.}
	\label{fig:ew_sim_obs}
\end{figure}

Figure \ref{fig:sbe_n2.5_halpha} shows no H$\alpha$ emission (blue dashed line) remains after dissipation, consistent with the diskless NRES/LCOGT 2020 observation. In Figure \ref{fig:ew_sim_obs}, the H$\alpha$ EW shows that this scenario (blue dashed line) requires $\sim17$ years to reproduce the diskless profile when $\alpha = 1$.

\begin{figure}
    \centering
	\makebox[\columnwidth][c]{\includegraphics[width = 1\columnwidth]{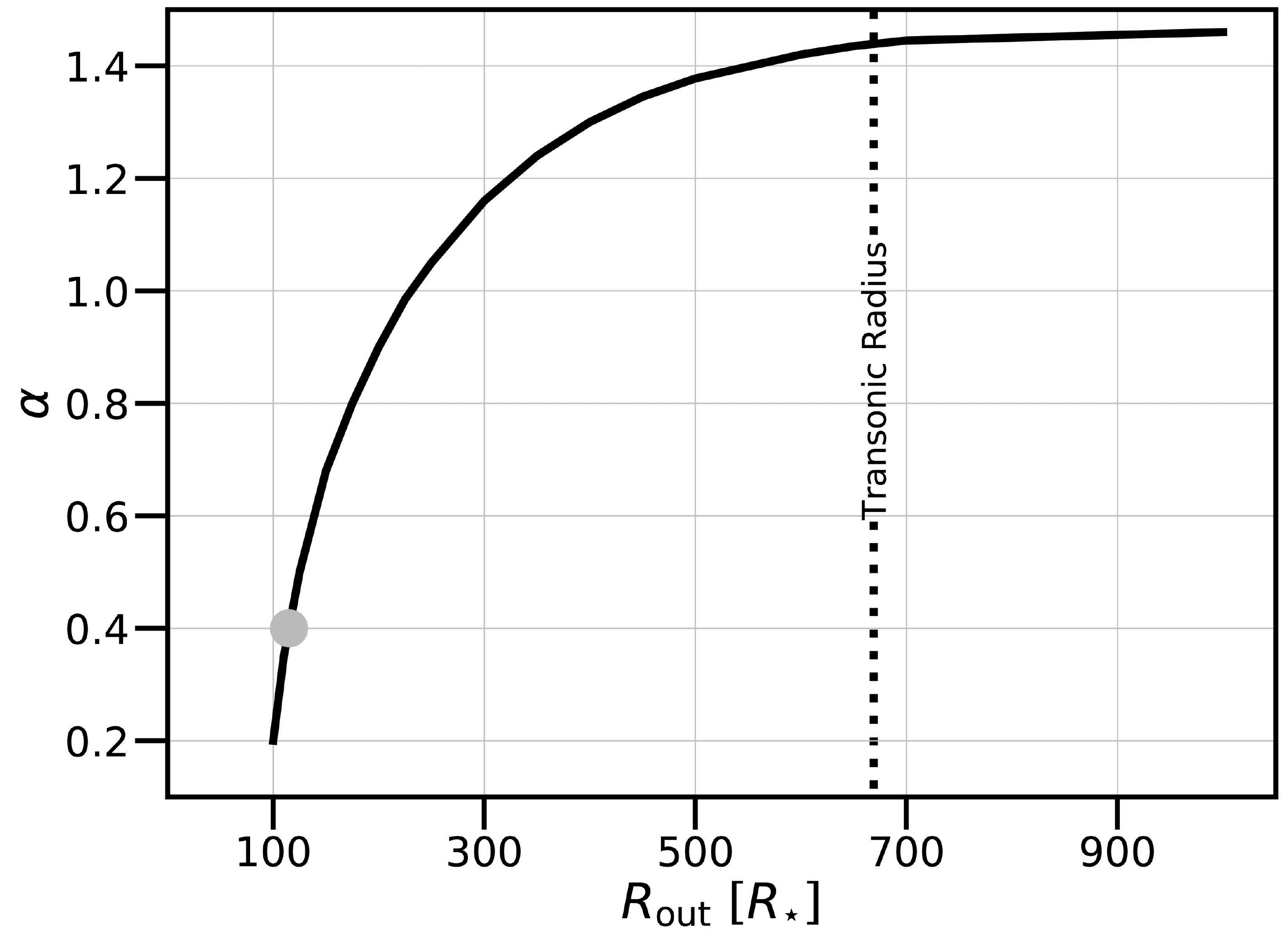}}
	\caption{Combinations of $\alpha$ and $R_{out}$ that allow the constant $\alpha T$, $n = 2.6$ scenario to lose all H$\alpha$ emission in 21 years. The grid of models has been interpolated. The vertical dotted line indicates the transonic radius \citep{okaz2001}. The models which best match the observed H$\alpha$ emission line dissipation are indicated by the point at $\alpha = 0.4$ and $R_{out} = 115~\rm{R_{eq}}$.}
	\label{fig:alpha_vs_rout}
\end{figure}

\begin{figure}
    \centering
	\makebox[\columnwidth][c]{\includegraphics[width = 1\columnwidth]{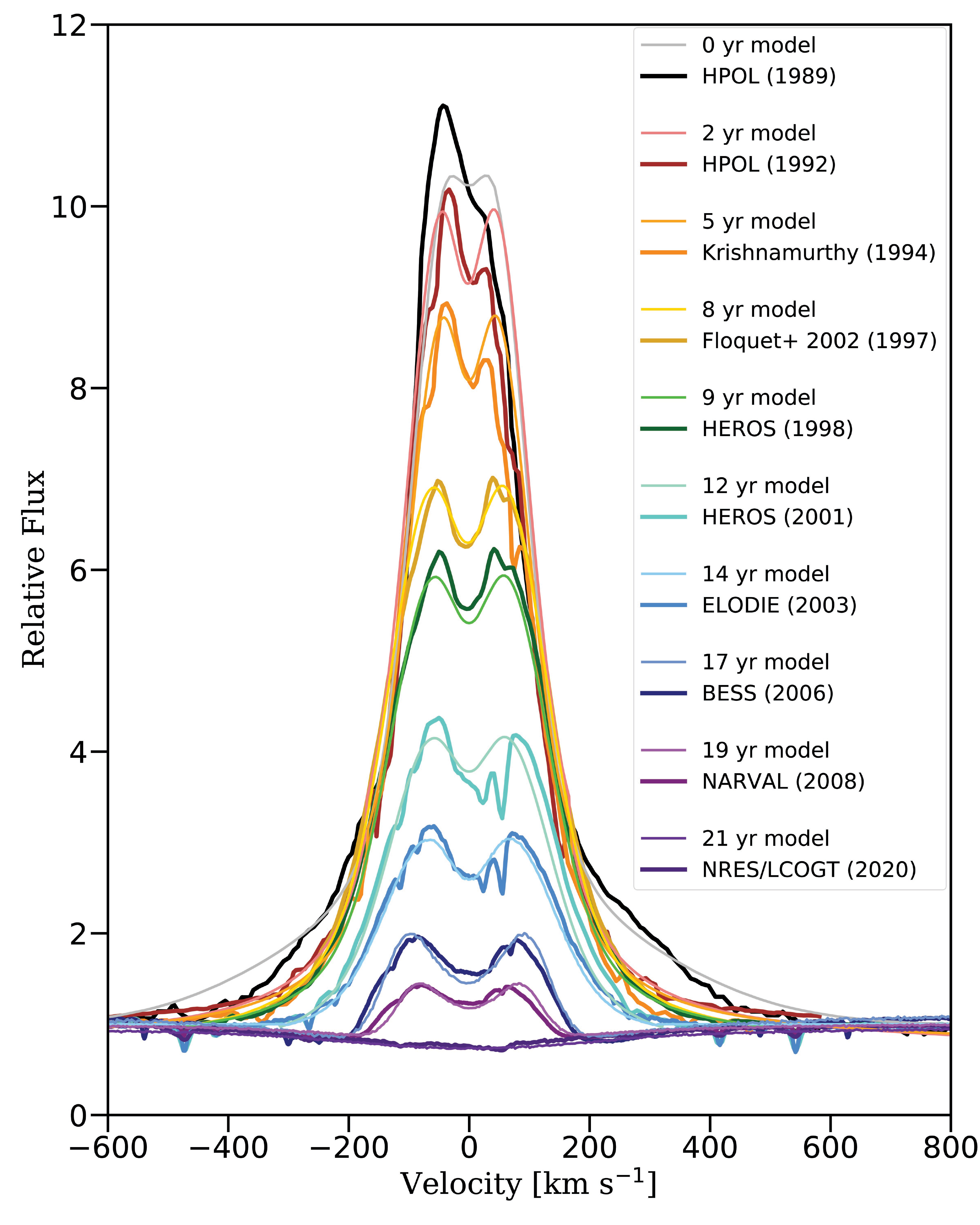}}
	\caption{H$\alpha$ emission line dissipation modelled with the constant $\alpha T$, $n = 2.6$ scenario. The thick, dark lines correspond to observations. The thinner, lighter lines are models using constant $\alpha$ and isothermal disks with $\alpha = 0.4$ and $R_{out} = 115~\rm{R_{eq}}$. The $0~\rm{year}$ model corresponds to ${\rho_0 = 2.5\times10^{-11}~{\rm g~cm}^{-3}}$ and ${n = 2.6}$.}
	\label{fig:obs_and_models}
\end{figure}

\begin{figure}
	\centering
	\makebox[\columnwidth][c]{\includegraphics[width = 1\columnwidth]{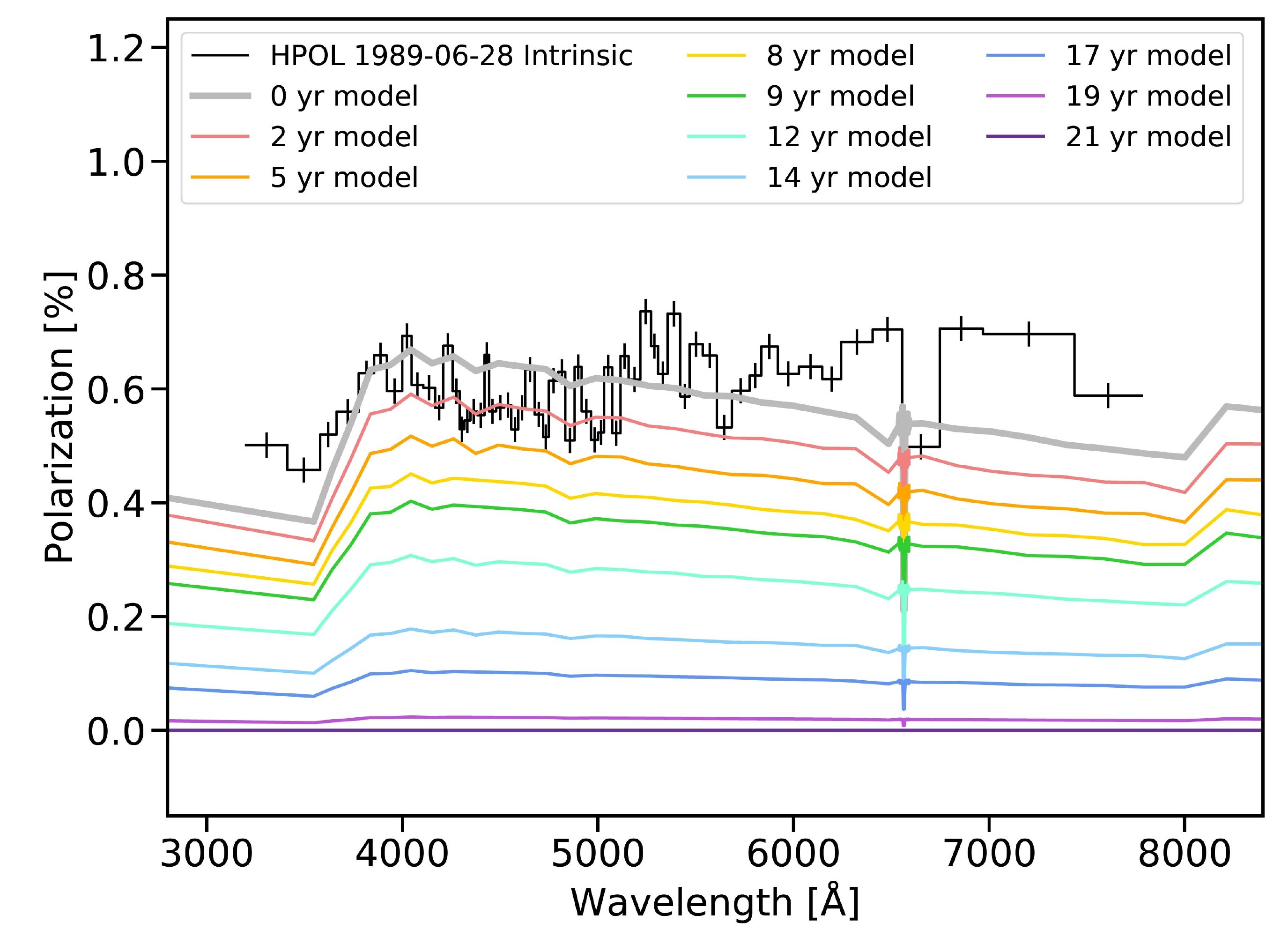}}
	\caption{Comparison of the polarization spectrum observed by HPOL on June 28th, 1989 and the models presented in Figure \ref{fig:sim_plotall}. The interstellar polarization, as determined in Section \ref{sec:observations}, was subtracted from the HPOL spectrum.}
	\label{fig:pol_serkowski}
\end{figure}

\begin{figure*}
	\centering
	\makebox[\textwidth][c]{\includegraphics[width = 1\textwidth]{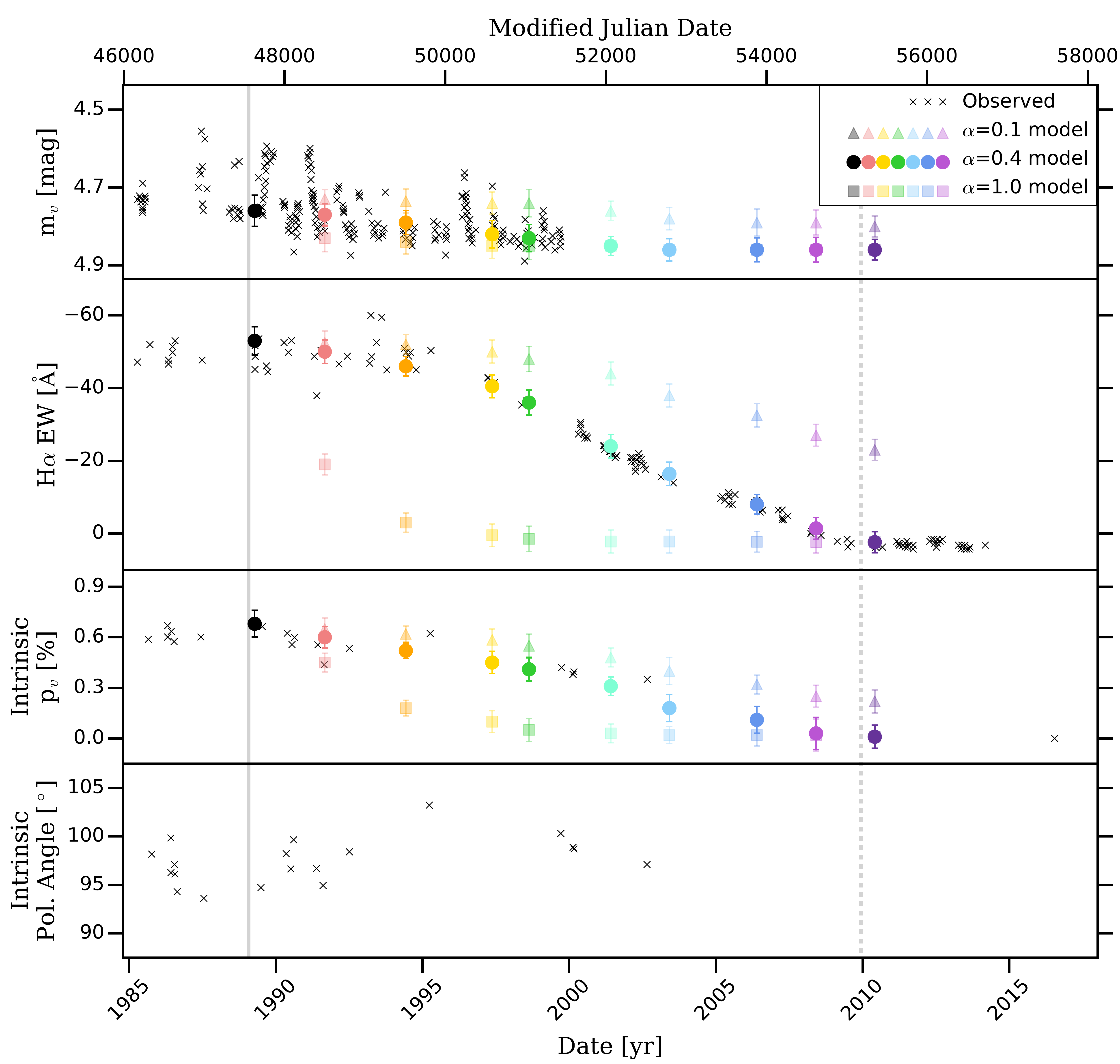}}
	\caption{Comparison of the observables shown in Figure \ref{fig:allplots} (black crosses) with the models presented in Figure \ref{fig:obs_and_models} (coloured circles) over the period of disk dissipation. Additional simulated observables for $\alpha = 0.1$ and $\alpha = 1$, and the observed polarization position angle intrinsic to the disk, are presented for comparison. The error bars show the 1$\sigma$ deviation of 10 simulations computed for each model. The color of each model corresponds with the same colors used in Figure \ref{fig:obs_and_models}. The vertical grey lines are the same as in Figure \ref{fig:allplots}.}
	\label{fig:sim_plotall}
\end{figure*}

Since $\alpha$ and $R_{out}$ both affect the rate of disk dissipation, they must be determined simultaneously. Figure~\ref{fig:alpha_vs_rout} shows an interpolated grid of models for the this scenario computed over the ranges of $0.2<\alpha<1.5$ and $100<R_{out}<1000~\rm{R_{eq}}$. The black line indicates which models dissipate from the initial state to having no H$\alpha$ emission in 21 years. The combination of $\alpha = 0.4$ and $R_{out} = 115~\rm{R_{eq}}$ was determined to best reproduce the observations acquired after $2$, $5$, $8$, $9$, $12$, $14$, $17$, $19$ and $21$ years of dissipation. Figure \ref{fig:ew_sim_obs} shows that this model closely reproduces the H$\alpha$ EW curve. This model is further compared to the data in Figures \ref{fig:obs_and_models}, \ref{fig:pol_serkowski}, and \ref{fig:sim_plotall}.

For each stage of dissipation in Figure \ref{fig:obs_and_models}, non-coherent electron scattering was accounted using the method described in Subsection \ref{subsec:results_1}. As previously mentioned, the $0~\rm{year}$ model required the electron thermal velocity in the disk to be $500~\rm{km~s^{-1}}$ when $22~\%$ of the light is scattered by the electrons. The electron thermal velocity required to broaden the line at all times during the dissipation is $\sim560\pm10~\rm{km~s^{-1}}$ (which corresponds to a kinetic temperature of $\sim12400\pm500~\rm{K}$). The fraction, $f$, of scattered light drops during dissipation, going from $f=20\%$ at $5~\rm{years}$ to $f=14\%$ at $12~\rm{years}$, and $f=4\%$ at $19~\rm{years}$. This is expected, as $f$ should decrease with decreasing disk density.

Figure \ref{fig:pol_serkowski} shows the observed polarization from HPOL near the onset of dissipation. The constant $\alpha T$, ${n = 2.6}$ scenario with $\alpha = 0.4$ and $R_{out} = 115~\rm{R_{eq}}$ is overplotted for comparison at the same time intervals as in Figure \ref{fig:obs_and_models}.

In Figure \ref{fig:sim_plotall}, we compare to each of the observables for $\alpha = 0.1,\, 0.4$ and $1.0$. The errors for each model were computed using a 1$\sigma$ deviation of ten unique simulations run for each model. The EW begins to change most rapidly nine years after the onset of dissipation, and decreases steadily for the following ten years until the line drops into absorption. For both the V-band magnitude and the continuum polarization the most rapid change occurs between zero and five years. During this time, the $\alpha = 0.4$ disk dims by $\sim0.1~\rm{mag}$. The percent polarization intrinsic to the $\alpha = 0.4$ disk drops steadily by $\sim0.2~\rm{\%}$ every five years. For comparison, the bottom panel shows the observed polarization position angle intrinsic to the disk, computed from the percent polarization. We find polarization position angles range between $\sim94^{\circ}$ and $\sim104^{\circ}$ and appear to remain constant through the dissipation event. These polarization position angles are consistent with the angle of $98^{\circ}$ determined in Section \ref{sec:observations}.

\subsubsection{Building and Dissipating a Non-Isothermal Disk}\label{subsubsec:results_2_3}

Here we investigate both the building and dissipation phases using the assumption of a non-isothermal disk. For reasons made apparent below, we refer to these models as the variable $\alpha T$, $n = 2.7$ scenario. In this case, the fluid equations no longer limit the density profile to $n = 3.5$, however we require knowledge of the now inseparable $\alpha T$ parameter. We assume a power-law relation between the product $\alpha T$ and the disk radius given by,

\begin{equation}
    \alpha(r)T(r) = \alpha_{0}T_{0}\bigg(\frac{R_{eq}}{r}\bigg)^{C}
    \label{eqn:alphapowerlaw}
\end{equation}

\noindent
where $\alpha(r)$ and $T(r)$ are the disk viscosity and disk temperature at radius $r$, $\alpha_{0}$ and $T_{0}$ are the disk viscosity and disk temperature at $r = R_{eq}$, and C is a constant power-law index. We chose $\alpha_{0} = 1$ and $T_{0} = 60\%$ of $T_{eff}$.

We tested values of $C$ ranging from 0.1 to 100 and found $C = 2$ to most closely match the $n = 2.6$ density slope. As shown in panel c) of Figure \ref{fig:sbe_triplot} the disk was unable to build to $n = 2.6$, with the majority of the outer disk having $n \approx 2.7$. The timescales required to build the outer disk to $n = 2.6$ were on the order of $1000$ years, significantly greater than time period that 66 Oph built its disk. The long viscous timescales at the outer disk are explained by the low viscosity inferred from Equation \ref{eqn:alphapowerlaw}. For instance, assuming a constant temperature and $C=2$, then $\alpha(r=100~\rm{R_{eq}}) / \alpha_0 = 0.0001$.

Figure \ref{fig:sbe_sed} shows that the SED for this scenario (yellow line) reproduces the observed visible spectrum and radio excess after $21$ years of dissipation, however it remains too bright at IR wavelengths. The H$\alpha$ line profile is shown in Figure \ref{fig:sbe_n2.5_halpha} (yellow line) after accounting for non-coherent electron scattering using the process outlined in Subsection~\ref{subsec:results_1}. We see that the emission remains too strong, evidence that the disk depletes too slowly, which is also seen in the H$\alpha$ EW in Figure \ref{fig:ew_sim_obs} (yellow line).

\subsubsection{Dissipating a Non-Isothermal Disk from a Defined Density}\label{subsubsec:results_2_4}

For our final dynamical scenario, we again adopt an ad hoc solution starting from the best quasi-steady model of Section \ref{subsubsec:results_2_1} to study the dissipation phase using the non-isothermal assumption. We refer to these models as the variable $\alpha T$, $n=2.6$ scenario. In this case, we assume that the outer disk could have built through past (undocumented) disk events, since Be stars remain on the main sequence for most of their lifetime \citep{geor2013}. We chose to use the $\alpha T$ profile described in Equation \ref{eqn:alphapowerlaw}, with $C = 2$.

Panel d) of Figure~\ref{fig:sbe_triplot} shows that the disk clears from the ``inside-out''. The most rapid change to the disk's density occurs within the first five years, beyond which the density decreases more slowly. The final five years of dissipation produces a change in density at $10~\rm{R_{eq}}$ of $\sim0.5$\%. As before, we find the viscous timescales of the outer disk are too large since $\alpha T$ decreases too rapidly with radius.

Figure \ref{fig:sbe_sed} shows this scenario (red line) does not match the diskless SED observations as closely as the isothermal models. After dissipation, this scenario more closely matches the diskless model from the UV to the near IR, and more closely matches the quasi-steady state in the far IR. Similarly to the variable $\alpha T$, $n = 2.7$ scenario, the excess radio flux remains bright, matching the observations.

Figure \ref{fig:sbe_n2.5_halpha} shows H$\alpha$ emission (red line) remains too strong after $21$ years of dissipation. Again, the line was adjusted to account for non-coherent electron scattering. During dissipation the peak flux of the line profile is reduced from $\sim$ eleven to nine times the continuum flux, but does not go into absorption. Dissipating for an additional $1000$ years, further decreases the peak flux of the line profile by a negligible $\sim0.1$\%. Figure \ref{fig:ew_sim_obs} shows the H$\alpha$ EW dissipates too slowly to match observations in this scenario (red line).

\section{Discussion and Summary} \label{sec:discussion}

The goal of this work is to use VDD theory to model the quasi-steady state of 66 Oph's disk, and then follow the subsequent $21$ year dissipation until the H$\alpha$ line transitions from emission to absorption. Using thousands of models constrained by photometric, polarimetric and H$\alpha$ observations, we determined the physical properties of the disk as it evolved.

We determined a set of stellar parameters (Table~\ref{tab:66ophpara}) using MCMC fitting of the UV spectrum. These parameters are consistent with those reported in the literature and were also confirmed by comparison to observations of 66 Oph obtained after dissipation (see Figure \ref{fig:sbe_sed}).

We used diskless polarimetric observations obtained in 2016 to determine the interstellar polarization in the direction of 66 Oph more accurately than previous studies using nearby field stars \citep[see the work by ][]{drap2014}. We modelled the interstellar polarization by fitting a Serkowski law to these diskless observations. The interstellar polarization was then subtracted from HPOL observations, acquired when the disk was present, to determine the polarization level and position angle of the disk (Figure \ref{fig:sim_plotall}). Using the QU diagram (Figure \ref{fig:quplot}) we also determined the disk's polarization position angle to be $\theta_{int} = 98\pm3\rm{^{\circ}}$. Our two methods of determining the polarization position angle were found to agree. This confirms our interstellar polarization correction using the 2016 LNA/OPD observations.

In this investigation, we begin by studying in detail the quasi-steady state phase of the disk, prior to the dissipation that started in 1989. Then, we used four different hydrodynamic scenarios to explore disk evolution. These scenarios assumed the disk was either isothermal or non-isothermal, and built to the quasi-steady state before dissipating or dissipated from a defined density distribution.

Our best fit disk model to the quasi-steady state, constrained by observations of the H$\alpha$ line profile and SED, and verified by the polarization, has a density of ${\rho_0 = 2.5\times10^{-11}~{\rm g~cm}^{-3}}$, ${n = 2.6}$, and $i = 57.5^{\circ}$ with the outer disk radius at the lower-limit of $R_{out} = 100~\rm{R_{eq}}$. The density slope, $n$, is consistent with the results of \citet{wate1987} who reported ${n = 2.5}$ and \citet{viei2017} who found ${n = 2.6}$, however, both of these studies found disk base densities at the innermost disk approximately one order of magnitude less dense. The close agreement of the density slope suggests it has been well constrained, however further study into the base density is necessary.

The scenario assuming an isothermal disk that builds and then dissipates (the constant $\alpha T$, $n=3.5$ scenario) reached a density slope of only $n = 3.5$, much steeper than the best fit quasi-steady state model with $n = 2.6$. As a result, the disk was not bright enough and dissipated too quickly due to the rapid density fall-off with radial distance. Interestingly, this scenario predicted a brightening of  the H$\alpha$ flux shortly after the onset of dissipation. We attribute this to the heating of cool regions of the disk at greater radial distance from the star as the innermost disk accretes. This phenomenon was not observed in disks with smaller values of $n$.

Next, we explored isothermal disks that dissipated from our best fit quasi-steady state (the constant $\alpha T$, $n=2.6$ scenario). We were able to constrain $R_{out}$ simultaneously with $\alpha$ as shown in Figure \ref{fig:alpha_vs_rout}. We find $R_{out} = 115~\rm{R_{eq}}$ and $\alpha = 0.4$. This scenario and these values successfully reproduced the rate of dissipation for all observables considered (H$\alpha$, V-band polarization and magnitude). A remarkable fit of H$\alpha$ line profile was obtained at selected epochs of disk dissipation (Figure \ref{fig:obs_and_models}).

\citet{krti2011} estimated the outer radius of Be star disks. Using their equation 5 and estimating the rate of change in the moment of inertia using the prescription from \citet{clar1989} along with our determination of $\dot{M}$ from \singlebe, we obtain an outer disk radius of $\sim125~\rm{R_{eq}}$, which is in relatively good agreement with our value determined above.

For the scenario with a non-isothermal disk that builds and dissipates (the variable $\alpha T$, $n=2.7$ scenario) we use a power-law $\alpha T$ that varies with radius as $r^{-2}$. The viscous timescales of this scenario was found to be too large to reproduce the rate of dissipation. After $21$ years of dissipation much of the disk remained, and while the visible and near IR flux match the diskless star, the H$\alpha$ and far IR remained too bright. However, the outer disk dissipated very little, and the predicted radio emission closely matched the JVLA observations from 2010.

Finally, we considered a non-isothermal disk and followed dissipation again from our best fit quasi-steady state (the variable $\alpha T$, $n=2.6$ scenario). We use the same procedure as the constant $\alpha T$, $n=2.6$ scenario, and the power-law $\alpha T$ from the variable $\alpha T$, $n=2.7$ scenario. This scenario dissipated similar to the other variable $\alpha T$ scenario, and the predicted observables reveal that the disk was still present after $21$ years.

Our models show that the excess radio emission observed in 2010 (Figure \ref{fig:sed_merged}) can only be reproduced if 66 Oph's outermost disk is unaffected by the dissipation. This suggests that a large mass reservoir was built in the outer disk, as discussed in \citet{rimu2018} and \citet{ghor2018}.

Overall, we find that the constant $\alpha T$, $n=2.6$ (isothermal) scenario, using $\alpha = 0.4$ and $R_{out} = 115~\rm{R_{eq}}$, best reproduces the observed dissipation of 66 Oph's disk. However, this scenario fails to explain how the disk could build to $n=2.6$. These smaller values of $n$ can be reached with a non-isothermal assumption, however all of our non-isothermal scenarios failed to match the timescale of dissipation. It is possible that other prescriptions for $\alpha T$, other than a power-law, might help this issue.

Another explanation for low values of $n$ might due to the accumulation effect described by \citet{pano2016} and by \citet{cyr2017}. Basically, the disk density becomes shallower due to the build up of material inward of a binary companion. This effect was not explored here, but we point out that our results indicate that the disk around 66 Oph is quite large, suggesting that the existence of a close by companion is unlikely.

In future work, we aim to better understand the structures of $\alpha$ and $T$ in the disk. It may be possible to empirically model $\alpha$ as a function of radius using a set of other time-series emission lines during the dissipation phase, as different lines form within different radial locations in the disk.

\acknowledgments{
The authors would like to thank the anonymous referee for careful reading and helpful comments that have greatly improved this paper. C. E. Jones wishes to acknowledge support though NSERC, the Natural Sciences and Engineering Research Council of Canada. This work was made possible through the use of the Shared Hierarchical Academic Research Computing Network (SHARCNET). A. C. Carciofi acknowledges support from CNPq (grant 311446/2019-1) and FAPESP (grant 2018/04055-8). Development and testing of the routines used was completed on the Laboratory of Astroinformatics (IAG/USP, NAT/Unicsul), whose purchase was made possible by the Brazilian agency FAPESP (grant 2009/54006-4) and the INCT-A. A. C. Rubio acknowledges the support of FAPESP grants 2017/08001-7 and 2018/13285-7. This work makes use of observations from the LCOGT network. K. C. Marr would like to thank Daniel Bednarski and Matheus Genaro from the Instituto de Astronomia, Geofísica e Ciências Atmosféricas, at the Universidade de São Paulo, for reducing the H$\alpha$ and polariation observations made at OPD, and guidance in working with the \emcee\ code, respectively. The authors would also like to thank Thomas Rivinius from the European Southern Observatory for providing the set of hydrogen spectral line observations used in this work, Chris Tycner from Central Michigan University for a set of H$\alpha$ equivalent width measurements, and Geraldine Peters for her generosity and willingness to share a set of H$\alpha$ observations.}

\software{\emcee\ \citep{fore2013},\ \hdust\ \citep{carc2006},\ \singlebe\ \citep{okaz2007}}

\bibliography{main}
\end{document}